%% file: sr_causal_dsi_v2.tex
\documentclass[onecolumn,10pt]{IEEEtran}
\input{preamble}

\usepackage{cite}
\usepackage{CJK}
\usepackage[ colorlinks = true,
             linkcolor = blue,
             urlcolor  = blue,
             citecolor = red,
             anchorcolor = green,]{hyperref}
\begin{document}

\title{Exponential Strong Converse for Successive Refinement with Causal Decoder Side Information}
\author{\IEEEauthorblockN{Lin Zhou and Alfred Hero
}\\
\IEEEauthorblockA{Department of EECS\\ University of Michigan\\
Emails: linzhou@umich.edu; hero@eecs.umich.edu}}
\maketitle

\begin{abstract}
We consider the $k$-user successive refinement problem with causal decoder side information and derive an exponential strong converse theorem. The rate-distortion region for the problem can be derived as a straightforward extension of the two-user case by Maor and Merhav (2008). We show that for any rate-distortion tuple outside the rate-distortion region of the $k$-user successive refinement problem with causal decoder side information, the joint excess-distortion probability approaches one exponentially fast. Our proof follows by judiciously adapting the recently proposed strong converse technique by Oohama using the information spectrum method, the variational form of the rate-distortion region and H\"older's inequality. The lossy source coding problem with causal decoder side information considered by El Gamal and Weissman is a special case ($k=1$) of the current problem. Therefore, the exponential strong converse theorem for the El Gamal and Weissman problem follows as a corollary of our result.
\end{abstract}

\section{Introduction}
We consider the $k$-user successive refinement problem with causal decoder side information shown in Figure~\ref{systemmodel}, which we refer to as the $k$-user causal successive refinement problem. The decoders aim to recover the source sequence based on the encoded symbols and causally available private side information sequences. Specifically, given the source sequence $X^n$, each encoder $f_j$ where $j\in\{1,\ldots,k\}$ compresses $X^n$ into a codeword $S_j$. At time $i\in\{1,\ldots,n\}$, for each $j\in\{1,\ldots,k\}$, the $j$-th user aims to recover the $i$-th source symbol using the codewords from encoders $(f_1,\ldots,f_j)$, the side information up to time $i$ and a decoding function $\phi_{j,i}$, i.e., $\hatX_{j,i}=\phi_{j,i}(S_1,\ldots,S_j,Y_{j,1},\ldots,Y_{j,i})$. Finally, at time $n$, for all $j\in\{1,\ldots,k\}$, the $j$-th user outputs the source estimate $\hatX_j^n$ which, under a distortion measure $d_j$, is required to be less than or equal to a specified distortion level $D_j$.

This problem was first considered by Maor and Merhav in \cite{maor2008} who fully characterized the rate-distortion region for the two-user version. Maor and Merhav showed that, unlike the case with non-causal side information~\cite{tian2007multistage}, no special structure e.g., degradedness, is required between the side information $Y_1^n$ and $Y_2^n$. However, Maor and Merhav only presented a \emph{weak} converse in \cite{maor2008}. In this paper, we strengthen the result in \cite{maor2008} by providing an exponential strong converse theorem for the $k$-user causal successive refinement problem, which states that the joint excess-distortion probability approaches one exponentially fast if the rate-distortion tuple falls outside the rate-distortion region.

\subsection{Related Works}
\label{sec:literature}
We first briefly summarize existing works on the successive refinement problem. The successive refinement problem was first considered by Equitz and Cover~\cite{equitz1991successive} and by Koshelev~\cite{koshelev1981estimation} who considered necessary and sufficient conditions for a source-distortion triple to be successively refinable. Rimoldi~\cite{rimoldi1994} fully characterized the rate-distortion region of the successive refinement problem under the joint excess-distortion probability criterion while Kanlis and Narayan~\cite{kanlis1996error} derived the excess-distortion exponent in the same setting. The second-order asymptotic analysis of No and Weissman~\cite{no2016}, which provides approximations to finite blocklength performance and implies strong converse theorems, was derived under the marginal excess-distortion probabilities criteria. This analysis was extended to the joint excess-distortion probability criterion by Zhou, Tan and Motani~\cite{zhou2016second}. Other frameworks for successive refinement decoding include \cite{tuncel2003additive,chowberger,tuncel2003,effros1999}.

The study of source coding with causal decoder side information was initiated by Weissman and El Gamal in \cite{Weissman2006causaltit} where they derived the rate-distortion function for the lossy source coding problem with causal side information at the decoders (i.e., $k=1$, see also \cite[Chapter 11.2]{el2011network}). Subsequently, Timo and Vellambi~\cite{timo2009twocausal} characterized the rate-distortion regions of the Gu-Effros two-hop network~\cite{gu2005source} and the Gray-Wyner problem~\cite{gray1974source} with causal decoder side information; Maor and Merhav~\cite{maor2010} derived the rate-distortion region for the successive refinement of the Heegard-Berger problem~\cite{heegard1985} with causal side information available at the decoders; Chia and Weissman~\cite{chia2011cascade} considered the cascade and triangular source coding problem with causal decoder side information. However, to the best of our knowledge, {\emph no} strong converse theorems exist for these problems.

As the information spectrum method will be used in this paper to derive an exponential strong converse theorem for the causal successive refinement problem, we briefly summarize the previous applications of this method to network information theory problems. In \cite{oohama2015wak,oohama2016new,oohama2018exponential}, Oohama used this method to derive exponential strong converses for the lossless source coding problem  with one-helper~\cite{ahlswede1975,wyner1975} (i.e., the Wyner-Ahlswede-K\"orner (WAK) problem), the asymmetric broadcast channel problem~\cite{korner1977abc}, and the Wyner-Ziv problem~\cite{wyner1976rate} respectively. Furthermore, Oohama's information spectrum method was also used to derive exponential strong converse theorems for content identification with lossy recovery~\cite{tuncel2014idenlossy} by Zhou, Tan, Yu and Motani~\cite{zhou2016cilossy} and for Wyner's common information problem under the total variation distance measure~\cite{wyner1975ci} by Yu and Tan~\cite{yu2017wyner}.

\subsection{Main Contribution and Challenges}
We consider the $k$-user causal successive refinement problem and present an exponential strong converse theorem. For given rates and blocklength, define the joint excess-distortion probability as the probability that either decoder incurs a distortion level greater than the specified distortion level (see \eqref{def:excessp}) and define the non-excess-distortion probability as the probability that all decoders satisfy the specified distortion levels (see \eqref{def:pcn}). Our proof proceeds as follows. First, we derive a non-asymptotic converse (finite blocklength upper) bound on the non-excess-distortion probability of any code for the $k$-user causal successive refinement problem using the information spectrum method. Subsequently, by using Cram\'er's inequality and the variational formulation of the rate-distortion region, we show that the non-excess-distortion probability decays exponentially fast to zero as the blocklength tends to infinity if the rate-distortion tuple falls outside the rate-distortion region of the causal successive refinement problem.

As far as we are aware, this paper is the first to establish a strong converse theorem for any lossy source coding problem with causal decoder side information. Furthermore, our methods can be used to derive exponential strong converse theorems for other lossy source coding problems with causal decoder side information discussed in Section \ref{sec:literature}. In particular, since the lossy source coding problems with causal decoder side information in~\cite{Weissman2006causaltit,maor2008} are special cases of the $k$-user causal successive refinement problem, the exponential strong converse theorems for the problems in~\cite{Weissman2006causaltit,maor2008} follow as a corollary of our result.

In order to establish the strong converse in this paper, we must  overcome several major technical challenges. The main difficulty lies in the fact that for the causal successive refinement problem, the side information is available to the decoder {\em causally} instead of non-causally. This causal nature of the side information makes the design of the decoder much more complicated and involved, which complicates the analysis of the excess-distortion probability. We find that classical strong converse techniques like the image size characterization~\cite{csiszar2011information} and the perturbation approach~\cite{wei2009strong} cannot lead to a strong converse theorem due to the above-mentioned difficulty. However, it is possible that other approaches different from ours can be used to obtain a strong converse theorem for the current problem. For example, it is interesting to explore whether two recently proposed strong converse techniques in~\cite{liubeyond,tyagi2018strong} can be used for this purpose considering the fact that the methods in~\cite{liubeyond,tyagi2018strong} have been successfully applied to problems including the Wyner-Ziv problem~\cite{wyner1976rate} and the Wyner-Ahlswede-K\"orner (WAK) problem~\cite{ahlswede1975,wyner1975}.

\section{Problem Formulation and Existing Results}
\subsection*{Notation}
Random variables and their realizations are in upper (e.g.,~$X$) and lower case (e.g.,\ $x$) respectively. Sets are denoted in calligraphic font (e.g.,\ $\mathcal{X}$). We use $\calX^{\mathrm{c}}$ to denote the complement of $\calX$ and use $X^n:=(X_1,\ldots,X_n)$ to denote a random vector of length $n$. Furthermore, given any $j\in[n]$, we use $X^{n\setminus j}$ to denote $(X_1,\ldots,X_{j-1},X_{j+1},\ldots,X^n)$.  We use $\mathbb{R}_+$ and $\bbN$ to denote the set of positive real numbers and integers respectively. Given two integers $a$ and $b$, we use $[a:b]$ to denote the set of all integers between $a$ and $b$ and use $[a]$ to denote $[1:a]$.  The set of all probability distributions on $\calX$ is denoted as $\calP(\calX)$ and the set of all conditional probability distributions from $\calX$ to $\calY$ is denoted as $\calP(\calY|\calX)$. For information-theoretic quantities such as entropy and mutual information, we follow the notation in \cite{csiszar2011information}. In particular, when the joint distribution of $(X,Y)$ is $P_{XY}\in\calP(\calX\times\calY)$, we use $I(P_X,P_{Y|X})$ and $I(X;Y)$ interchangeably.

\subsection{Problem Formulation}
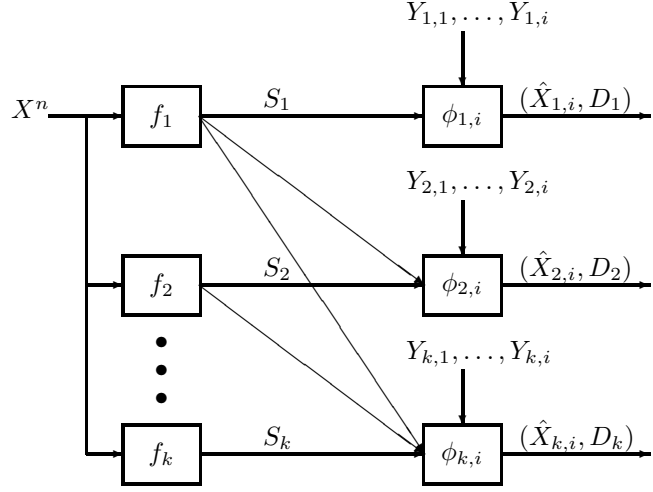
\begin{figure}[t]
\centering
\setlength{\unitlength}{0.5cm}
\scalebox{1}{
\begin{picture}(18,12)
\linethickness{1pt}
\put(3,9){\framebox(2,1.5){$f_1$}}
\put(1,9.75){\vector(1,0){2}}
\put(0,9.7){\makebox{$X^n$}}
\put(6.7,10){\makebox{$S_1$}}
\put(5,9.75){\vector(1,0){6}}
\put(10.5,12.3){\makebox{$Y_{1,1},\ldots,Y_{1,i}$}}
\put(11,9){\framebox(2,1.5){$\phi_{1,i}$}}
\put(12,12){\vector(0,-1){1.5}}
\put(13,9.75){\vector(1,0){4}}
\put(13.5,10){\makebox{$(\hatX_{1,i},D_1)$}}

\put(2,9.75){\line(0,-1){4.5}}
\put(2,5.25){\vector(1,0){1}}
\put(3,4.5){\framebox(2,1.5){$f_2$}}
\put(6.7,5.5){\makebox{$S_2$}}
\put(5,5.25){\vector(1,0){6}}
\put(11,4.5){\framebox(2,1.5){$\phi_{2,i}$}}
\put(12,7.5){\vector(0,-1){1.5}}
\put(10.5,7.8){\makebox{$Y_{2,1},\ldots,Y_{2,i}$}}
\put(13.5,5.5){\makebox{$(\hatX_{2,i},D_2)$}}
\put(13,5.25){\vector(1,0){4}}
\put(5,9.75){\vector(4,-3){6}}
\put(4,3.75){\circle*{0.3}}
\put(4,3){\circle*{0.3}}
\put(4,2.25){\circle*{0.3}}
\put(2,5.25){\line(0,-1){4.5}}
\put(2,0.75){\vector(1,0){1}}
\put(3,0){\framebox(2,1.5){$f_k$}}
\put(6.7,1){\makebox{$S_k$}}
\put(5,0.75){\vector(1,0){6}}
\put(11,0){\framebox(2,1.5){$\phi_{k,i}$}}
\put(12,3){\vector(0,-1){1.5}}
\put(10.5,3.3){\makebox{$Y_{k,1},\ldots,Y_{k,i}$}}
\put(13.5,1){\makebox{$(\hatX_{k,i},D_k)$}}
\put(13,0.75){\vector(1,0){4}}
\put(5,5.25){\vector(4,-3){6}}
\put(5,9.75){\vector(2,-3){6}}
\end{picture}}
\caption{Encoder-decoder system model for the $k$-user successive refinement problem with causal decoder side information at time $i\in[n]$. Each encoder $f_j$ where $j\in[k]$ compresses the source information into codewords $S_j$. Given accumulated side information $(Y_{j,1},\ldots,Y_{j,i})$ and the codewords $(S_1,\ldots,S_j)$, decoder $\phi_{j,i}$ reproduces the $i$-th source symbol as $\hatX_{j,i}$. At time $n$, for $j\in[k]$, the estimate $\hatX_j^n$ for user $j$ is required to satisfy distortion constraint $D_j$ under a distortion measure $d_j$.}
\label{systemmodel}
\end{figure}

Let $k\in\bbN$ be a fixed finite integer and let $P_{XY^k}$ be a joint probability mass function (pmf) on the finite alphabet $\calX\times(\prod_{j\in[k]}\calY_j)$ with its marginals denoted in the customary way, e.g., $P_X$, $P_{XY_1}$. Throughout the paper, we consider memoryless sources $(X^n,Y_1^n,\ldots,Y_k^n)$, which are generated i.i.d. according to $P_{XY^k}$. Let $\hat{\calX}_j$ be the alphabet of the reproduced source symbol for user $j\in[k]$. Recall the encoder-decoder system model for the $k$-user causal successive refinement problem in Figure \ref{systemmodel}.

A formal definition of a code for the causal successive refinement problem is as follows.
\begin{definition}
An $(n,M_1,\ldots,M_k)$-code for the causal successive refinement problem consists of
\begin{itemize}
\item $k$ encoding functions
\begin{align}
f_j:\calX^n\to\calM_j:=\{1,\ldots,M_j\},~j\in[k],
\end{align}
\item and $kn$ decoding functions: for each $i\in[n]$
\begin{align}
\phi_{j,i}:&(\prod_{l\in[j]}\calM_l)\times(\calY_j)^i\to\hat{\calX}_j,~j\in[k].
\end{align}
\end{itemize}
\end{definition}

For $j\in[k]$, let $d_j:\calX\times\hat{\calX}_j\to[0,\infty)$ be a distortion measure. Given the source sequence $x^n$ and a reproduced version $\hatx_j^n$, we measure the distortion between them using the additive distortion measure $d_j(x^n,\hatx_j^n):=\frac{1}{n}\sum_{i\in[n]}d_j(x_i,\hatx_{j,i})$. To evaluate the performance of an $(n,M_1,\ldots,M_k)$-code for the causal successive refinement problem, given distortion specified levels $(D_1,\ldots,D_k)$, we consider the following joint excess-distortion probability
\begin{align}
\rmP_\rme^{(n)}(D_1,\ldots,D_k):=\Pr\big\{\exists~j\in[k]~\mathrm{s.t.~}d_j(X^n,\hatX_j^n)>D_j\big\}\label{def:excessp}.
\end{align}
For ease of notation, throughout the paper, we use $D^k$ to denote $(D_1,\ldots,D_k)$, $M^k$ to denote $(M_1,\ldots,M_k)$ and $R^k$ to denote $(R_1,\ldots,R_k)$.

Given $\varepsilon\in(0,1)$, the $\varepsilon$-rate-distortion region for the $k$-user causal successive refinement problem is defined as follows.

\begin{definition}
\label{epsregion}
Given any $\varepsilon\in(0,1)$, a rate-distortion tuple $(R^k,D^k)$ is said to be $\varepsilon$-achievable if there exists a sequence of $(n,M^k)$-codes such that
\begin{align}
\limsup_{n\to\infty}\frac{1}{n}\log M_1&\leq R_1\label{r1},\\
\limsup_{n\to\infty}\frac{1}{n}\log M_j&\leq R_j-\sum_{l\in[j-1]}R_l,~\forall~j\in[2:k],\label{r2}\\
\limsup_{n\to\infty} \rmP_\rme^{(n)}(D^k)&\leq \varepsilon\label{jd:criterion}.
\end{align}
The closure of the set of all $\varepsilon$-achievable rate-distortion tuples is called the $\varepsilon$-rate-distortion region and is denoted as $\calR(\varepsilon)$.
\end{definition}
Note that in Definition \ref{epsregion}, $R_j$ is the sum rate of the first $j$ decoders. Using Definition \ref{epsregion}, the rate-distortion region for the problem is defined as
\begin{align}
\calR&:=\bigcap_{\varepsilon\in(0,1)}\calR(\varepsilon)\label{def:calr}.
\end{align}

\subsection{Existing Results}
For the two-user causal successive refinement problem, the rate-distortion region was fully characterized by Maor and Merhav~\cite[Theorem 1]{maor2008}. With slight generalization, the result can be extended to $k$-user case.

For $j\in[k]$, let $W_j$ be a random variable taking values in a finite alphabet $\calW_j$. For simplicity, throughout the paper, we let
\begin{align}
T&:=(X,Y^k,W^k,\hatX^k)\label{def:T4ease},
\end{align}
and let $(t,\calT)$ be a particular realization of $T$ and its alphabet set, respectively.

Define the following set of joint distributions:
\begin{align}
\calP^*
\nn&:=\Big\{Q_T\in\calP(\calT): Q_{XY^k}=P_{XY^k},~W^k-X-Y^k,~|\calW_1|\leq |\calX|+3,~\mathrm{and~}\forall~j\in[k]:\\*
&\qquad\qquad|\calW_j|\leq |\calX|\Big(\prod_{l\in[j-1]}|\calW_l|\Big)+1,~\hatX_j=\phi_j(W^j,Y_j)~\mathrm{for~some~}\phi_j:\Big(\prod_{l\in[j]}\calW_l\Big)\times\calY_j\to\hat{\calX}_j\Big\}\label{def:calp*}.
\end{align}
Given any joint distribution $Q_T\in\calP(\calT)$, define the following set of rate-distortion tuples
\begin{align}
\calR(Q_T)
\nn:=\Big\{(R^k,D^k):~R_1&\geq I(Q_X,Q_{W_1|X}), D_1\geq \mathbb{E}[d_1(X,\phi_1(W_1,Y_1))],
~\mathrm{and~}\forall~j\in[2:k]:\\
R_j-\sum_{l\in[j-1]}R_l&\geq I(Q_{X|W^{j-1}},Q_{W_j|XW^{j-1}}|Q_{W_{j-1}}),~D_j\geq \mathbb{E}[d_j(X,\phi_j(W^j,Y_j))]\Big\}\label{def:calrqt}.
\end{align}
Maor and Merhav~\cite{maor2008} defined the following information theoretical sets of rate-distortion tuples
\begin{align}
\calR^*:=\bigcup_{Q_T\in\calP^*} \calR(Q_T)\label{def:calr*}.
\end{align}
\begin{theorem}
\label{causalsrweakconverse}
The rate-distortion region for the causal successive refinement problem satisfies
\begin{align}
\calR=\calR^*.
\end{align}
\end{theorem}
We remark that in \cite{maor2008}, Maor and Merhav considered the average distortion criterion for $k=2$, i.e.,
\begin{align}
\limsup_{n\to\infty}\mathbb{E}[d_j(X^n,\hatX_j^n)]\leq D_k,~\forall~j\in[k],\label{ad:criterion}
\end{align} 
instead of the vanishing joint excess-distortion probability criterion (see \eqref{jd:criterion}) in Definition \ref{epsregion}. However, with slight modification to the proof of \cite{maor2008}, it can be verified (see Appendix \ref{proof:causalweak}) that the rate-distortion region $\calR$ under the vanishing joint excess-distortion probability criterion, is identical to the rate-distortion region $\calR^*$ derived by Maor and Merhav under the average distortion criterion.

Theorem \ref{causalsrweakconverse} implies that if a rate-distortion tuple falls outside the rate-distortion region, i.e., $(R^k,D^k)\notin\calR$, then the excess-distortion probability $\rmP_\rme^{(n)}(D^k)$ is bounded away from zero. We strengthen the converse proof of Theorem \ref{causalsrweakconverse} by showing that if $(R^k,D^k)\notin\calR$, then the excess-distortion probability $\rmP_\rme^{(n)}(D^k)$ approaches one exponentially fast as the blocklength $n$ tends to infinity.

\section{Main Results}

\subsection{Preliminaries}
In this subsection, we present necessary definitions and a key lemma before stating our main result.

Define the following set of distributions
\begin{align}
\calQ
&:=\big\{Q_T\in\calP(\calT): |\calW_j|\leq \big(|\calX||\calY||\calZ||\hat{\calX}_1||\hat{\calX_2}|\big)^j,~\forall~j\in[k]\big\}\label{def:calQ}.
\end{align}
Given any $(\mu,\alpha^k,\beta^k)\in\bbR_+\times [0,1]^{2k}$ such that 
\begin{align}
\sum_{i\in[k]}(\alpha_i+\beta_i)=1\label{linearconstraints},
\end{align}
for any $Q_T\in\calQ$, define the following linear combination of log likelihoods
\begin{align}
\omega_{Q_T}^{(\mu,\alpha^k,\beta^k)}(t)
\nn&:=\log\frac{Q_X(x)}{P_X(x)}+\log\frac{Q_{Y^k|XW^k}(y^k|x,w^k)}{P_{Y^k|X}(y^k|x)}+
\log\frac{Q_{XY^{k\setminus 1}W^{k\setminus 1}|Y_1W_1\hatX_1}(x,y^{k\setminus 1},w^{k\setminus 1}|y_1,w_1,\hatx_1)}{Q_{XY^{k\setminus 1}W^{k\setminus 1}|Y_1W_1}(x,y^{k\setminus 1},w^{k\setminus 1}|y_1,w_1)}\\*
\nn&\qquad+\sum_{j\in[2:k]}\log\frac{Q_{\hatX_j|XY^kW^k\hatX^{j-1}}(\hatx_j|x,y^k,w^k,\hatx^{j-1})}{Q_{\hatX_j|Y_jW^j}(\hatx_j|y_j,w^j)}+\mu\alpha_1\log\frac{Q_{X|W_1}(x|w_1)}{P_X(x)}\\*
&\qquad+\sum_{j\in[2:k]}\mu\alpha_j\log\frac{Q_{X|W^j}(x|w^j)}{Q_{X|W^{j-1}}(x|w^{j-1})}+\sum_{j\in[k]}\mu\beta_j d_j(x,\hatx_j).\label{def:omegaabg}
\end{align}
Given any $\theta\in\bbR_+$ and any $Q_T\in\calQ$, define the minus cumulant generating function of $\omega_{Q_T}^{(\mu,\alpha^k,\beta^k)}(\cdot)$ as
\begin{align}
\Omega^{(\theta,\mu,\alpha^k,\beta^k)}(Q_T)
&:=-\log \mathbb{E}_{Q_T}\big[\exp\big(-\theta \omega_{Q_T}^{(\mu,\alpha^k,\beta^k)}(T)\big)\big].\label{def:OmegaQT}
\end{align}
Furthermore, define the minimal minus cumulant generating function over distributions in $\calQ$ as
\begin{align}
\Omega^{(\theta,\mu,\alpha^k,\beta^k)}
&:=\min_{Q_T\in\calQ}\Omega^{(\theta,\mu,\alpha^k,\beta^k)}(Q_T).\label{def:Omega}
\end{align}

Finally, given any rate-distortion tuple $(R^k,D^k)$, define
\begin{align}
\kappa^{(\alpha^k,\beta^k)}(R^k,D^k)&:=\alpha_1R_1+\beta_1D_1+\sum_{j\in[2:k]}(\alpha_j (R_j-\sum_{l\in[j-1]}R_l)+\beta_j D_j)\label{def:kappamuab}\\
F^{(\theta,\mu,\alpha^k,\beta^k)}(R^k,D^k)
&:=\frac{\Omega^{(\theta,\mu,\alpha^k,\beta^k)}-\theta\mu\kappa^{(\alpha^k,\beta^k)}(R^k,D^k)}{1+(2k+2)\theta+\sum_{j\in[k]}2\theta\mu\alpha_j},\label{def:Ftabgm}\\
F(R^k,D^k)
&:=\sup_{(\theta,\mu,\alpha^k,\beta^k)\in\bbR_+^2\times[0,1]^{2k}:~\eqref{linearconstraints}}F^{(\theta,\mu,\alpha^k,\beta^k)}(R^k,D^k)\label{def:F}.
\end{align}

With the above definitions, we have the following lemma establishing the properties of the exponent function $F(R^k,D^k)$.
\begin{lemma}
\label{prop:F}
The following claims hold.
\begin{enumerate}
\item For any rate-distortion tuple outside the rate-distortion region, i.e., $(R^k,D^k)\notin\calR$, we have
\begin{align}
F(R^k,D^k)>0,
\end{align}
\item For any rate-distortion tuple inside the rate-distortion region, i.e., $(R^k,D^k)\in\calR$, we have
\begin{align}
F(R^k,D^k)=0.
\end{align}
\end{enumerate}
\end{lemma}
The proof of Lemma \ref{prop:F} is inspired by \cite[Property 4]{oohama2018exponential}, \cite[Lemma 2]{zhou2016cilossy} and is given in Section \ref{proof:propF}. As will be shown in Theorem \ref{theorem:main}, the exponent function $F(R^k,D^k)$ is a lower bound on the exponent of the probability of non-excess-distortion probability for the $k$-user causal successive refinement problem. Thus, Claim (i) in Lemma \ref{prop:F} is crucial to establish the exponential strong converse theorem which states that the excess-distortion probability (see \eqref{def:excessp}) approaches one exponentially fast with respect to the blocklength of the source sequences.

\subsection{Main Result}
Define the probability of non-excess-distortion as
\begin{align}
\rmP_\rmc^{(n)}(D^k)
&:=1-\rmP_\rme^{(n)}(D^k)=\Pr\big\{\forall~j\in[k],~d_j(X^n,\hatX_j^n)\leq D_j\big\}\label{def:pcn}.
\end{align}

\begin{theorem}
\label{theorem:main}
Given any $(n,M^k)$-code for the $k$-user causal successive refinement problem such that
\begin{align}
\log M_1\leq nR_1,\mathrm{and~}\forall~j\in[2:k],~\log M_j\leq n(R_j-\sum_{l\in[j-1]}R_l)\label{conditions},
\end{align}
we have the following non-asymptotic upper bound on the probability of non-excess-distortion
\begin{align}
\rmP_\rmc^{(n)}(D^k)
&\leq (2k+3)\exp(-nF(R^k,D^k))\label{upp_cd}.
\end{align}
\end{theorem}
The proof of Theorem \ref{theorem:main} is given in Section \ref{sec:proofmain}. Several remarks are in order.

First, our result is non-asymptotic, i.e., the bound in \eqref{upp_cd} holds for any $n\in\bbN$. In order to prove Theorem \ref{theorem:main}, we adapt the recently proposed strong converse technique by Oohama~\cite{oohama2018exponential} to analyze the probability of non-excess-distortion probability. We first obtain a non-asymptotic upper bound using the information spectrum of log-likelihoods involved in the definition of $\omega_{Q_T}^{(\mu,\alpha^k,\beta^k)}$ (see \eqref{def:omegaabg}) and then apply Cram\'er's bound on large deviations (see e.g., \cite[Lemma 13]{zhou2016cilossy}) to obtain an exponential type non-asymptotic upper bound. Subsequently, we apply the recursive method~\cite{oohama2018exponential} and proceed similarly as in \cite{zhou2016cilossy} to obtain the desired result. Our method can also be used to establish similar results for other source coding problems with causal decoder side information~\cite{timo2009twocausal,chia2011cascade,maor2010}.

Second, we believe that classical strong converse techniques including the image size characterization~\cite{csiszar2011information} and the perturbation approach~\cite{wei2009strong} cannot lead to the strong converse theorem for the causal successive refinement problem. The main obstacle is that the side information is available \emph{causally} and thus complicates the decoding analysis significantly.

Invoking Lemma \ref{prop:F} and Theorem \ref{theorem:main}, we conclude that the exponent on the right hand side of \eqref{upp_cd} is positive if and only if the rate-distortion tuple is outside the rate-distortion region, which implies the following exponential strong converse theorem.

\begin{theorem}
\label{expstrongconverse}
For any sequence of $(n,M^k)$-codes satisfying the rate constraints in \eqref{conditions}, given any distortion levels $D^k$, we have that if $(R^k,D^k)\notin\calR$, then the probability of correct decoding $\rmP_\rmc^{(n)}(D^k)$ decays exponentially fast to zero as the blocklength of the source sequences tends to infinity.
\end{theorem}

As a result of Theorem \ref{expstrongconverse}, we conclude that for every $\varepsilon\in(0,1)$, the $\varepsilon$-rate distortion region (see Definition \ref{epsregion}) satisfies that
\begin{align}
\calR(\varepsilon)=\calR\label{strongconverse},
\end{align}
i.e., strong converse holds for the $k$-user causal successive refinement problem. Using the strong converse theorem and Marton's change-of-measure technique~\cite{Marton74}, similarly to \cite[Theorem 5]{zhou2016cilossy}, we can also derive an upper bound on the exponent of the excess-distortion probability. Furthermore, applying the one-shot techniques in \cite{yassaee2013technique}, we can also establish a non-asymptotic achievability bound. Applying the Berry-Esseen theorem to the achievability bound and analyzing the non-asymptotic converse bound in Theorem \ref{theorem:main}, similarly to \cite{oohama2018exponential}, we conclude that the backoff from the rate-distortion region at finite blocklength scales on the order of $\Theta(\frac{1}{\sqrt{n}})$. However, nailing down the exact second-order asymptotics~\cite{kostina2012fixed,TanBook} is challenging and is left for future work.

Our main results in Lemma \ref{prop:F}, Theorems \ref{theorem:main} and \ref{expstrongconverse} can be specialized to the settings in \cite{Weissman2006causaltit,maor2008} with $k=1$ and $k=2$ respectively.

\section{Proof of the Non-Asymptotic Converse Bound (Theorem \ref{theorem:main})}
\label{sec:proofmain}

\subsection{Preliminaries}
Given any $(n,M^k)$-code with encoding functions $(f_1,\ldots,f_k)$ and and decoding functions $\{(\phi_{1,i},\ldots,\phi_{k,i})\}_{i\in[n]}$, we define the following induced conditional distributions: for each $j\in[k]$,
\begin{align}
P_{S_j|X^n}(s_j|x^n)
&:=1\{s_j=f_j(x^n)\},\\
P_{\hatX_j^n|S^jY_j^n}(\hatx_j^n|s^j,y^n)
&:=\prod_{i\in[n]}1\{\hatx_{j,i}=\phi_{j,i}(s^j,y^i)\}\label{def:softdecoding}.
\end{align}
For simplicity, in the following, we let
\begin{align}
G&:=(X^n,Y_1^n,\ldots,Y_k^n,S^k,\hatX_1^n,\ldots,\hatX_k^n),\label{def:G4ease}
\end{align}
and let $(g,\calG)$ be a particular realization and the alphabet of $G$ respectively. With above definitions, we have that the distribution $P_G$ satisfies that for any $g\in\calG$,
\begin{align}
P_G(g)
&:=P_{XY^k}^n(x^n,y_1^n,\ldots,y_k^n)
\big(\prod_{j\in[k]}P_{S_j|X^n}(s_j|x^n)\big)\big(\prod_{j\in[k]}P_{\hatX_j^n|S^jY_j^n}(\hatx_j^n|s^j,y^n)\big)\label{def:joint}.
\end{align}
In the remaining part of this section, all distributions denoted by $P$ are induced by the joint distribution $P_G$.

For simplicity, given any $(i,j)\in[n]\times[k]$, we use $Y_{j,1}^{j,i}$ to denote $(Y_{j,1},\ldots,Y_{j,i})$ and we use $Y_{1,i}^{k,i}$ to denote $(Y_{1,i},\ldots,Y_{k,i})$. Similarly, we use $W_{1,i}^{k,i}$ and $\hatX_{1,i}^{k,i}$. For each $i\in[n]$, let auxiliary random variables be $W_{1,i}:=(X^{i-1},Y_{1,1}^{1,i-1},\ldots,Y_{k,1}^{k,i-1},S_1)$ and $W_{j,i}=S_j$ for all $j\in[2:k]$. Note that as a function of $i\in[n]$, the Markov chain $(W_{1,i}^{k,i})-X_i-(Y_i,Z_i)$ holds under $P_G$. Throughout the paper, for each $i\in[n]$, we let
\begin{align}
T_i&:=(X_i,Y_{1,i}^{k,i},W_{1,i}^{k,i},\hatX_{1,i}^{k,i})\label{def:Ti4ease},
\end{align} 
and let $(t_i,\calT_i)$ be a particular realization and the alphabet of $T_i$, respectively.

For each $i\in[n]$, let $Q_{C_i|D_i}$ be arbitrary distributions where $C_i\in\calT_i$ and $D_i\in\calT_i$. Given any positive real number $\eta$, define the following subsets of $\calG$:
\begin{align}
\calB_1&:=
\Big\{g:0\geq \frac{1}{n}\sum_{i\in[n]}\log\frac{Q_{X_i}(x_i)}{P_X(x_i)}-\eta\Big\}\label{def:calB1},\\
\calB_2&:=
\Big\{g:0\geq \frac{1}{n}\sum_{i\in[n]}\log\frac{Q_{Y_{1,i}^{k,i}|X_iW_{1,i}^{k,i}}(y_{1,i}^{k,i}|x_i,w_{1,i}^{k,i})}{P_{Y^k|X}(y_{1,i}^{k,i}|x_i)}-\eta\Big\}\label{def:calB2},\\
\calB_3
&:=\bigg\{
g:0\geq \frac{1}{n}\sum_{i\in[n]}\log\frac{Q_{X_iY_{2,i}^{k,i}W_{2,i}^{k,i}|Y_{1,i}W_{1,i}\hatX_{1,i}}(x_i,y_{2,i}^{k,i},w_{2,i}^{k,i}|y_{1,i},w_{1,i},\hatx_{1,i})}{P_{X_iY_{2,i}^{k,i}W_{2,i}^{k,i}|Y_{1,i}W_{1,i}}(x_i,y_{2,i}^{k,i},w_{2,i}^{k,i}|y_{1,i},w_{1,i})}-\eta\bigg\}\label{def:calB3},\\
\calB_4&:=\bigg\{
g:0\geq \frac{1}{n}\sum_{i\in[n]}\log\frac{Q_{\hatX_{j,i}|X_iY_{1,i}^{k,i}W_{1,i}^{k,i}\hatX_{1,i}^{j-1,i}}(\hatx_{j,i}|x_i,y_{1,i}^{k,i},w_{1,i}^{k,i},\hatx_{1,i}^{j-1,i})}{P_{\hatX_{j,i}|Y_{j,i}W_{1,i}^{j,i}}(\hatx_{j,i}|y_{j,i},w_{1,i}^{j,i})}-\eta,~\forall~j\in[2:k]\bigg\}\label{def:calB4},\\
\calB_5&:=\bigg\{
g:R_1\geq \frac{1}{n}\sum_{i\in[n]}\log\frac{P_{X_i|W_{1,i}}(x_i|w_{1,i})}{P_X(x_i)}-\eta\bigg\}\label{def:calB5},\\
\calB_6&:=\bigg\{
g:R_j-\sum_{l\in[j-1]}R_l\geq \frac{1}{n}\sum_{i\in[n]}\log\frac{P_{X_i|W_{1,i}^{j,i}}(x_i|w_{1,i}^{j,i})}{P_{X_i|W_{1,i}^{j-1,i}}(x_i|w_{1,i}^{j-1,i})}-\eta,~\forall~j\in[2:k]\bigg\}\label{def:calB6}\\
\calB_7&:=\bigg\{
g:D_j\geq \frac{1}{n}\sum_{i\in[n]}\log\exp(d_j(x_i,\hatx_{j,i})),~\forall~j\in[k]\bigg\}\label{def:calB8}.
\end{align}

\subsection{Proof Steps}
\begin{lemma}
\label{fbl-sl}
For any $(n,M^k)$-code satisfying \eqref{conditions}, given any distortion levels $D^k$, we have
\begin{align}
\rmP_\rmc^{(n)}(D^k)
\leq \Pr\Big\{\bigcap_{i\in[7]}\calB_i\Big\}+(2k+2)\exp(-n\eta).
\end{align}
\end{lemma}
The proof of Lemma \ref{fbl-sl} is given in Appendix \ref{proof:fbl-sl} and divided into two steps. First, we derive a $n$-letter non-asymptotic upper bound which holds for certain arbitrary $n$-letter auxiliary distributions. Subsequently, we single-letterize the derived bound by proper choice of auxiliary distributions and careful decomposition of induced distributions of $P_G$.

For simplicity, in the following, we will use $P_i$ to denote $P_{T_i}$ and use $Q_i$ to denote $Q_{T_i}$. Given any $\mu\in\bbR_+$ and any $(\alpha^k,\beta^k)\in[0,1]^{2k}$ satisfying \eqref{linearconstraints}, define
\begin{align}
\nn f_{Q_i,P_i}^{(\alpha^k,\beta^k)}(t_i)
&:=\frac{Q_{X_i}(x_i)}{P_X(x_i)}\frac{Q_{Y_{1,i}^{k,i}|X_iW_{1,i}^{k,i}}(y_{1,i}^{k,i}|x_i,w_{1,i}^{k,i})}{P_{Y^k|X}(y_{1,i}^{k,i}|x_i)}\frac{Q_{X_iY_{2,i}^{k,i}W_{2,i}^{k,i}|Y_{1,i}W_{1,i}\hatX_{1,i}}(x_i,y_{2,i}^{k,i},w_{2,i}^{k,i}|y_{1,i},w_{1,i},\hatx_{1,i})}{P_{X_iY_{2,i}^{k,i}W_{2,i}^{k,i}|Y_{1,i}W_{1,i}}(x_i,y_{2,i}^{k,i},w_{2,i}^{k,i}|y_{1,i},w_{1,i})}\\*
\nn&\qquad\times \bigg(\prod_{j\in[2:k]}\log\frac{Q_{\hatX_{j,i}|X_iY_{1,i}^{k,i}W_{1,i}^{k,i}\hatX_i^{j-1}}(\hatx_{j,i}|x_i,y_{1,i}^{k,i},w_{1,i}^{k,i},\hatx_{1,i}^{j-1,i})}{P_{\hatX_{j,i}|Y_{j,i}W_{1,i}^{j,i}}(\hatx_{j,i}|y_{j,i},w_{1,i}^{j,i})}\bigg)\frac{P^{\mu\alpha_1}_{X_i|W_{1,i}}(x_i|w_{1,i})}{P_X^{\mu\alpha_1}(x_i)}\\*
&\qquad\times \bigg(\prod_{j\in[2:k]}\frac{P^{\mu\alpha_j}_{X_i|W_{1,i}^{j,i}}(x_i|w_{1,i}^{j,i})}{P^{\mu\alpha_j}_{X_i|W_{1,i}^{j-1,i}}(x_i|w_{1,i}^{j-1,i})}\bigg)\exp\Big(\mu \big(\sum_{j\in[k]}\beta_j d_j(x_i,\hatx_{j,i})\big)\Big)\label{def:fQPimabg}.
\end{align}
Furthermore, given any non-negative real number $\lambda\in\bbR_+$, define
\begin{align}
\Omega^{(\lambda,\mu,\alpha^k,\beta^k)}(\{P_i,Q_i\}_{i\in[n]})
&:=-\log\mathbb{E}\Big[\exp\big(-\lambda \sum_{i\in[n]}\log f_{Q_i,P_i}^{(\mu,\alpha^k,\beta^k)}(T_i)\big)\Big]\label{def:OmegaQiPi}.
\end{align}

Recall the definition of $\kappa^{(\alpha^k,\beta^k)}(R^k,D^k)$ in \eqref{def:kappamuab}. Using Cram\'er's bound~\cite[Lemma 13]{zhou2016cilossy}, we obtain the following non-asymptotic exponential type upper bound on the probability of non-excess-distortion, whose proof is given in in Appendix \ref{proof:fbletype}.
\begin{lemma}
\label{fbletype}
For any $(n,M^k)$-code satisfying the conditions in Lemma \ref{fbl-sl}, given any distortion levels $D^k$, we have
\begin{align}
\rmP_\rmc^{(n)}(D^k)
&\leq (2k+3)\exp\bigg(-n\frac{\frac{1}{n}\Omega^{(\lambda,\mu,\alpha^k,\beta^k)}(\{P_i,Q_i\}_{i\in[n]})-\lambda\mu\kappa^{(\alpha^k,\beta^k)}(R^k,D^k)}{1+\lambda(k+2+\sum_{j\in[k]}\mu\alpha_j)}\bigg).
\end{align}
\end{lemma}

Furthermore, let
\begin{align}
\underline{\Omega}^{(\lambda,\mu,\alpha^k,\beta^k)}(\{P_i\}_{i\in[n]}
&:=\inf_{n\in\bbN}\sup_{\{Q_i\}_{i\in[n]}}\Omega^{(\lambda,\mu,\alpha^k,\beta^k)}(\{P_i,Q_i\}_{i\in[n]}\label{def:uOmegaPi}.
\end{align}
Given any $(\lambda,\mu,\alpha^k)\in\bbR_+^2\times[0,1]^k$ such that 
\begin{align}
\lambda(k+\sum_{j\in[k]}\mu\alpha_j)\leq 1,\label{constraintlambda}
\end{align}
let 
\begin{align}
\theta:=\frac{\lambda}{1-k\lambda-\sum_{j\in[k]}\lambda\mu\alpha_j}\label{def:theta}.
\end{align}
then we have
\begin{align}
\lambda=\frac{\theta}{1+k\theta+\sum_{j\in[k]}\theta\mu\alpha_j}\label{lambda=f(theta)}.
\end{align}

The following lemma which relates $\underline{\Omega}^{(\lambda,\mu,\alpha^k,\beta^k)}(\{P_i\}_{i\in[n]}$ with $\Omega^{(\theta,\mu,\alpha^k,\beta^k)}$ (recall \eqref{def:Omega}) is crucial.
\begin{lemma}
\label{vital}
Given any $(\lambda,\mu,\alpha^k,\beta^k)\in\bbR_+^2\times[0,1]^3$ satisfying \eqref{linearconstraints} and \eqref{constraintlambda}, we have that for $\theta$ defined in \eqref{def:theta},
\begin{align}
\underline{\Omega}^{(\lambda,\mu,\alpha^k,\beta^k)}(\{P_i\}_{i\in[n]})
\geq \frac{n\Omega^{(\theta,\mu,\alpha^k,\beta^k)}}{1+k\theta+\sum_{j\in[k]}\theta\mu\alpha_j}.
\end{align}
\end{lemma}
The proof of Lemma \ref{vital} is given in Appendix \ref{proof:vital}. In the proof of Lemma \ref{vital}, we apply H\"older's inequality and the recursive method used in \cite{oohama2018exponential}.

Combining Lemmas \ref{fbletype} and \ref{vital}, we conclude that for any $(n,M^k)$-code satisfying the conditions in Lemma \ref{fbl-sl} and for any $(\mu,\alpha^k,\beta^k)\in\bbR_+\times[0,1]^3$, given any $\lambda\in\bbR_+$ satisfying \eqref{constraintlambda}, we have
\begin{align}
\rmP_\rmc^{(n)}(D^k)
&\leq 7\exp\bigg(-n\frac{\frac{1}{n}\Omega^{(\lambda,\mu,\alpha^k,\beta^k)}(\{P_i,Q_i\}_{i\in[n]})-\lambda\mu\kappa^{(\alpha^k,\beta^k)}(R^k,D^k)}{1+\lambda(k+2+\sum_{j\in[k]}\mu\alpha_j)}\bigg)\\
&\leq 7\exp\bigg(-n\frac{\Omega^{(\theta,\mu,\alpha^k,\beta^k)}-\theta\mu\kappa^{(\alpha^k,\beta^k)}(R^k,D^k)}{1+(2k+2)\theta+\sum_{j\in[k]}2\theta\mu\alpha_j}\bigg)\label{usedefinitionskappa}\\
&\leq 7\exp\big(-nF^{(\theta,\mu,\alpha^k,\beta^k)}(R^k,D^k)\big)\label{usededefF},
\end{align}
where \eqref{usedefinitionskappa} follows from the definitions of $\kappa^{(\alpha^k,\beta^k)}(\cdot)$ in \eqref{def:kappamuab}, $\theta$ in \eqref{def:theta} and the results in \eqref{lambda=f(theta)}, and \eqref{usededefF} follows from the definition of $F^{(\theta,\mu,\alpha^k,\beta^k)}(\cdot)$ in \eqref{def:F}.

\section{Proof of Properties of Strong Converse Exponent: Proof of Lemma \ref{prop:F}}
\label{proof:propF}
\subsection{Alternative Expressions of the Rate-Distortion Region}

In this section, we present several definitions and an alternative characterization of the rate-distortion region $\calR$ using the supporting hyperplanes, which facilitate the proof of Lemma \ref{prop:F}.

Define the following set of joint distributions
\begin{align}
\calP
\nn&:=\big\{Q_T\in\calP(\calT): Q_{XY^k}=P_{XY^k},~W^k-X-Y^k,~\mathrm{and}~\forall~j\in[k]:\\*
&\qquad\qquad |\calW_j|\leq|\calX|(\prod_{l\in[j-1]}|\calW_l|)+1,~\hatX_j-(W^j,Y_j)-(X,Y^{k\setminus j},W_{j+1}^k,\hatX^{j-1})\big\}\label{def:calp}.
\end{align}
Recall the definition of $\calR(Q_T)$ in \eqref{def:calrqt}. Define
\begin{align}
\calR_{\rm{ran}}
&:=\bigcup_{Q_T\in\calP}\calR(Q_T).
\end{align}
Furthermore, let $\calP_{\rm{sh}}$ be the following set of joint distributions
\begin{align}
\calP_{\rm{sh}}
&:\big\{Q_T\in\calP(\calT): Q_{XY^k}=P_{XY^k},~W^k-X-Y^k,~\mathrm{and}~\forall~j\in[k],\\*
&\qquad\qquad\qquad |\calW_j|\leq(|\calX|)^j,~\hatX_j-(W^j,Y_j)-(X,Y^{k\setminus j},W_{j+1}^k,\hatX^{j-1})\big\}\label{def:calpsh}.
\end{align}
Given any $(\alpha^k,\beta^k)\in[0,1]^{2k}$ satisfying \eqref{linearconstraints}, define the following linear combination of achievable rate-distortion tuples
\begin{align}
\rmR^{(\alpha^k,\beta^k)}
&:=\min_{Q_T\in\calP_{\rm{sh}}}
\big\{
\alpha_1 I(Q_X,Q_{W_1|X})+\sum_{j\in[2:k]}\alpha_j I(Q_{X|W^{j-1}},Q_{W_j|XW^{j-1}}|Q_{W^{j-1}})+\sum_{j\in[k]}\beta_j \mathbb{E}[d_j(X,\hatX_j)]
\big\}\label{def:Rabg}.
\end{align}
Recall the definition of $\kappa^{\cdot}(\cdot)$ in \eqref{def:kappamuab}. Finally, let $\calR_{\rm{sh}}$ be the following collection of rate-distortion tuples
\begin{align}
\calR_{\rm{sh}}
&:=\bigcap_{(\alpha^k,\beta^k)\in[0,1]^{2k}:~\eqref{linearconstraints}} \big\{(R^k,D^k):
\kappa^{(\alpha^k,\beta^k)}(R^k,D^k)\geq \rmR^{(\alpha^k,\beta^k)}\big\}\label{def:calrsh}.
\end{align}

Recall the definitions of $\calR$ in \eqref{def:calr} and $\calR^*$ in \eqref{def:calr*}. Similarly to \cite[Properties 2 and 3]{oohama2018exponential}, one can show that the rate-distortion region $\calR$ for the $k$-user causal successive refinement problem remains unchanged even if one uses stochastic decoding functions. Furthermore, the rate-distortion region $\calR$ has the following alternative characterization using supporting hyperplanes.
\begin{lemma}
\label{rdregion:alternative}
The rate-distortion region for the causal successive refinement problem satisfies
\begin{align}
\calR=\calR^*=\calR_{\rm{ran}}=\calR_{\rm{sh}}.
\end{align}
\end{lemma}

\subsection{Proof of Claim (i)}
Recall that we use $T$ (see \eqref{def:T4ease}) to denote the collection of random variables $(X,Y^k,S^k,\hatX^k)$ and use $t,\calT$ similarly to denote a realization of $T$ and its alphabet, respectively. For any $P_T\in\calP_{\rm{sh}}$ (recall \eqref{def:calpsh}), any $(\alpha^k,\beta^k)\in[0,1]^{2k}$ satisfying \eqref{linearconstraints} and any $\lambda\in\bbR_+$, for any $t\in\calT$, define
\begin{align}
\tilde{\omega}^{(\alpha^k,\beta^k)}_{P_T}(t)
&:=\alpha_1\log\frac{P_{X|W_1}(x|w_1)}{P_X(x)}+\sum_{j\in[2:k]}\alpha_j\log\frac{P_{X|W^j}(x|w^j)}{P_{X|W_{j-1}}(x|w_{j-1})}+\sum_{j\in[k]}\beta_j d_j(x,\hatx_j)\label{def:tildeomegatPT},\\
\tilde{\Omega}^{(\lambda,\alpha^k,\beta^k)}(P_T)
&:=-\log\mathbb{E}_{P_T}\big[\exp(-\lambda \tilde{\omega}^{(\alpha^k,\beta^k)}_{P_T}(T))\big]\label{def:tildeOmegaPT}.
\end{align}
For simplicity, we let
\begin{align}
\alpha^+:=\max_{j\in[k]}\alpha_j\label{def:alpha+}.
\end{align}
Furthermore, paralleling \eqref{def:Omega} to \eqref{def:F} and recalling \eqref{def:kappamuab}, let
\begin{align}
\tilde{\Omega}^{(\lambda,\alpha^k,\beta^k)}
&:=\min_{P_T\in\calP_{\rm{sh}}}\tilde{\Omega}^{(\lambda,\alpha^k,\beta^k)}(P_T)\label{def:tildeOmega},\\
\tilF^{(\lambda,\alpha^k,\beta^k)}(R^k,D^k)
&:=\frac{\tilde{\Omega}^{(\lambda,\alpha^k,\beta^k)}-\lambda\kappa^{(\alpha^k,\beta^k)}(R^k,D^k)}{2k+3+\lambda\alpha^+ +\sum_{j\in[2:k]}\lambda(2k+3)\alpha_j+\sum_{l\in[k]}2\lambda\alpha_l}\label{def:tilFlabg},\\
\tilF(R^k,D^k)
&:=\sup_{(\lambda,\alpha^k,\beta^k)\in\bbR_+\times[0,1]^{2k}:~\eqref{linearconstraints}}\tilF^{(\lambda,\alpha^k,\beta^k)}(R^k,D^k)\label{def:tilF}.
\end{align}
For subsequent analysis, define the following tilted distribution
\begin{align}
P_T^{(\lambda,\alpha^k,\beta^k)}(t)
&:=\frac{P_T(t)\exp(-\lambda \tilde{\omega}^{(\alpha^k,\beta^k)}_{P_T}(t))}{\mathbb{E}_{P_T}\big[\exp(-\lambda \tilde{\omega}^{(\alpha^k,\beta^k)}_{P_T}(T))\big]}\label{def:tiltPT}.
\end{align}
Finally, define the following dispersion function
\begin{align}
\rho&:=\sup_{P_T\in\calP_{\rm{sh}}}\sup_{(\lambda,\alpha^k,\beta^k)\in\bbR_+\times[0,1]^{2k}:~\eqref{linearconstraints}}\mathrm{Var}_{P_T^{(\lambda,\alpha^k,\beta^k)}}\big[\tilde{\omega}^{(\alpha^k,\beta^k)}_{P_T}(T)\big]\label{def:rho}.
\end{align}
Note that $\rho$ is positive and finite.

The proof of Claim (i) in Lemma \ref{prop:F} is completed by the following lemma which relates $F(R^k,D^k)$ with $\tilF(R^k,D^k)$.
\begin{lemma}
\label{relateF&tilF}
The following claims hold.
\begin{enumerate}
\item For any rate-distortion tuple $(R^k,D^k)$,
\begin{align}
F(R^k,D^k)\geq \tilF(R^k,D^k).
\end{align}
\item For any rate-distortion tuple $(R^k,D^k)$ outside the rate-distortion region, i.e., $(R^k,D^k)\notin\calR$, we have that for some $\delta\in(0,\rho]$,
\begin{align}
\tilF(R^k,D^k)\geq \frac{\delta^2}{2(2k+9)\rho}>0.
\end{align}
\end{enumerate}
\end{lemma}
The proof of Lemma \ref{relateF&tilF} is inspired by \cite{oohama2018exponential,zhou2016cilossy} and given in Appendix \ref{proof:relateF&tilF}. In order to prove Lemma \ref{relateF&tilF}, we use the alternative characterizations of the rate-distortion region $\calR$ in Lemma \ref{rdregion:alternative} and analyze the connections between the two exponent functions $F(R^k,D^k)$ and $\tilF(R^k,D^k)$.

\subsection{Proof of Claim (ii)}
If a rate-distortion tuple falls inside the rate-distortion region, i.e., $(R^k,D^k)\in\calR$, then there exists a distribution $Q_T^*\in\calP_{\rm{sh}}$ (see \eqref{def:calpsh}) such that for any $(\alpha^k,\beta^k)\in[0,1]^{2k}$ satisfying \eqref{linearconstraints}, we have
\begin{align}
\kappa^{(\alpha^k,\beta^k)}(R^k,D^k)
\nn&\geq \alpha_1 I(Q_{X_1}^*,Q_{W_1|X_1}^*)+\beta_1^*\mathbb{E}[d_1(X,\hatX_1)]\\*
&\qquad+\sum_{j\in[2:k]}(\alpha_j^* I(Q_{X_1|W^{j-1}}^*,Q_{W_j|XW^{j-1}}^*|Q_{W^{j-1}}^*)+\beta_j^*\mathbb{E}[d_j(X,\hatX_j)])\label{consequenceinside}.
\end{align}
Recall the definition of $\Omega^{(\theta,\mu,\alpha^k,\beta^k)}(Q_T)$ in \eqref{def:OmegaQT}. From simple calculation, we have that
\begin{align}
\Omega^{(\theta,\mu,\alpha^k,\beta^k)}(Q_T)&=0\label{zerovalue},\\
\frac{\partial \Omega^{(\theta,\mu,\alpha^k,\beta^k)}(Q_T)}{\partial \theta}\bigg|_{\theta=0}
&=\mathbb{E}_{Q_T}\big[\omega_{Q_T}^{\mu,\alpha^k,\beta^k}(T)\big],\label{firstdOmegaQT}\\
\frac{\partial \Omega^{(\theta,\mu,\alpha^k,\beta^k)}(Q_T)}{\partial \theta}
&<0\label{seconddOmegaQT}.
\end{align}
Combining \eqref{zerovalue} to \eqref{seconddOmegaQT}, by applying Taylor expansions, we have that for any $(\theta,\mu,\alpha^k,\beta^k)\in\bbR_+^2\times[0,1]^{2k}$,
\begin{align}
\Omega^{(\theta,\mu,\alpha^k,\beta^k)}(Q_T)
&\leq \theta\mathbb{E}_{Q_T}\big[\omega_{Q_T}^{\mu,\alpha^k,\beta^k}(T)\big]\label{resultitouse}.
\end{align}

Using the definition of $\Omega^{(\theta,\mu,\alpha^k,\beta^k)}$ in \eqref{def:Omega}, we conclude that
\begin{align}
\Omega^{(\theta,\mu,\alpha^k,\beta^k)}
&\leq\min_{Q_T\in\calP_{\rm{sh}}}\Omega^{(\theta,\mu,\alpha^k,\beta^k)}(Q_T)\label{calPincalQ}\\
&\leq \min_{Q_T\in\calP_{\rm{sh}}}\theta\mathbb{E}_{Q_T}\big[\omega_{Q_T}^{\mu,\alpha^k,\beta^k}(T)\big]\label{usetaylorupp}\\
\nn&\leq\alpha_1 I(Q_{X_1}^*,Q_{W_1|X_1}^*)+\beta_1^*\mathbb{E}[d_1(X,\hatX_1)]\\*
&\qquad+\sum_{j\in[2:k]}(\alpha_j^* I(Q_{X_1|W^{j-1}}^*,Q_{W_j|XW^{j-1}}^*|Q_{W^{j-1}}^*)+\beta_j^*\mathbb{E}[d_j(X,\hatX_j)])\label{usemarkovchain}\\
&\leq \mu\kappa^{(\alpha^k,\beta^k)}(R^k,D^k),\label{useconinside}
\end{align}
where \eqref{calPincalQ} follows since $\calP_{\rm{sh}}\subseteq\calQ$ (recall \eqref{def:calQ}), \eqref{usetaylorupp} follows from the result in \eqref{resultitouse}, \eqref{usemarkovchain} follows from the definitions of $\omega_{Q_T}^{\mu,\alpha^k,\beta^k}(t)$ in \eqref{def:OmegaQT} and $\calP_{\rm{sh}}$ in \eqref{def:calpsh}, and \eqref{useconinside} follows from the result in \eqref{consequenceinside}.

Using the definition of $F^{(\theta,\mu,\alpha^k,\beta^k)}(R^k,D^k)$ in \eqref{def:F} and the result in \eqref{useconinside}, we conclude that for any $(R^k,D^k)\in\calR$,
\begin{align}
F^{(\theta,\mu,\alpha^k,\beta^k)}(R^k,D^k)\leq 0.
\end{align}
The proof of Claim (ii) is completed by noting that
\begin{align}
\lim_{\theta\to 0}F^{(\theta,\mu,\alpha^k,\beta^k)}(R^k,D^k)=0.
\end{align}

\section{Conclusion}

We considered the $k$-user causal successive refinement problem~\cite{maor2008} and established an exponential strong converse theorem using the strong converse techniques proposed by Oohama~\cite{oohama2018exponential}. Our work appears to be the first to derive a strong converse theorem for any source coding problem with causal decoder side information. The methods we adopted can also be used to obtain exponential strong converse theorems for other source coding problems with causal decoder side information. This paper further illustrates the usefulness and generality of Oohama's information spectrum method in deriving exponential strong converse theorems. We believe that using Oohama's techniques~\cite{oohama2018exponential}, the strong converse theorem for channel coding with causal state information~\cite{shannon1958channels,sigurjonsson2005multiple} can also be established.

There are several natural future research directions. In Theorem \ref{theorem:main}, we presented only an lower bound on the strong converse exponent. It would be worthwhile to obtain an exact strong converse exponent and thus characterize the exact speed at which the probability of non-excess-distortion decays exponentially fast with respect to the blocklength of source sequences when the rate-distortion tuple falls outside the rate-distortion region. Furthermore, one can explore whether the methods in this paper can be used to establish strong converse theorems for causal successive refinement under the logarithmic loss~\cite{shkel2018single,courtade2014multiterminal}, which corresponds to soft decoding of each source symbol. Finally, one can also explore extensions to continuous alphabet by considering Gaussian memoryless sources under bounded distortion measures and derive second-order asymptotics~\cite{strassen1962asymptotische,hayashi2009information,polyanskiy2010finite,TanBook,kostina2013lossy} for the causal successive refinement problem.

\appendix

\subsection{Proof of Theorem \ref{causalsrweakconverse}}
\label{proof:causalweak}
Replacing \eqref{jd:criterion} with Definition \ref{epsregion}, we can define the $\varepsilon$-rate-distortion region $\calR_{\rm{ad}}(\varepsilon)$ under the average distortion criterion. Furthermore, let
\begin{align}
\calR_{\rm{ad}}&:=\bigcap_{\varepsilon\in[0,1)}\calR_{\rm{ad}}(\varepsilon).
\end{align}
Maor and Merhav~\cite{maor2008} showed that for $k=2$,
\begin{align}
\calR_{\rm{ad}}=\calR^*.
\end{align}
Actually, in \cite[Section VII]{maor2008}, in order to prove that $\calR^*\subseteq\calR_{\rm{ad}}$, it was already shown that $\calR^*\subseteq\calR$. Furthermore, it is straightforward to show that the above results hold for any finite $k\in\bbN$. Thus, to prove Theorem \ref{causalsrweakconverse}, it suffices to show
\begin{align}
\calR\subseteq \calR^*=\calR_{\rm{ad}}.
\end{align}
For this purpose, given any $j\in[k]$, let
\begin{align}
\bard_j&:=\max_{(x,\hatx_j)\calX\times\hat{\calX}_j}d_j(x,\hatx_j).
\end{align}
From the problem formulation, we know that $\bard_j<\infty$ for all $j\in[k]$. Now consider any rate-distortion tuple $(R^k,D^k)\in\calR$, then we have \eqref{r1} to \eqref{jd:criterion}. Therefore, for any $j\in[k]$,
\begin{align}
\limsup_{n\to\infty}\mathbb{E}[d_j(X^n,\hatX_j^n)]
&\leq \limsup_{n\to\infty}\Big(\mathbb{E}[d_j(X^n,\hatX_j^n)1\{d_j(X^n,\hatX_j^n)\leq D_j\}]+\bard_j\Pr\{d_j(X^n,\hatX_j^n)>D_j\}\Big)\\
&\leq D_j.
\end{align}
As a result, we have $(R^k,D^k)\in\calR_{\rm{ad}}$. Therefore, we have shown that $\calR\subseteq\calR_{\rm{ad}}=\calR^*$.

\subsection{Proof of Lemma \ref{fbl-sl}}
\label{proof:fbl-sl}
Recall the definition of $G$ and $\calG$ in \eqref{def:G4ease}. Given any $C\in\calG$ and $D\in\calG$, let $Q_{C|D}$ be arbitrary distributions. For simplicity, given each $j\in[k]$, we use $\bY^j$ to denote $(Y_1^n,\ldots,Y_j^n)$ and use $\bY^{j\setminus l}$ to denote $(Y_1^n,\ldots,Y_{l-1}^n,Y_{l+1}^n,\ldots,Y_l^n)$ where $l\in[j]$. Similarly we use $\hat{\bX}^j$ and $\hat{\bX}^{j\setminus l}$. 

Given any positive real number $\eta$, define the following sets:
\begin{align}
\calA_1&:=
\Big\{g:\frac{1}{n}\log\frac{P_X^n(x^n)}{Q_{X^n}(x^n)}\geq -\eta\Big\}\label{def:calA1},\\
\calA_2&:=
\Big\{g:\frac{1}{n}\log\frac{P_{Y^k|X}^n(\by^k|x^n)}{Q_{\bY^k|X^nS^k}(\by^k|x^n,s^k)}\geq -\eta\Big\}\label{def:calA2},\\
\calA_3&:=\Big\{g:\frac{1}{n}\log\frac{P_{X^n\bY^{k\setminus 1}S^{k\setminus 1}|Y_1^nS_1}(x^n,\by^{k\setminus 1},s^{k\setminus 1}|y_1^n,s_1)}{Q_{X^n\bY^{k\setminus 1}S^{k\setminus 1}|Y_1^nS_1\hatX_1^n}(x^n,\by^{k\setminus 1},s^{k\setminus 1}|y_1^n,s_1,\hatx_1^n)}\geq -\eta\Big\}\label{def:calA3},\\
\calA_4&:=
\Big\{g:\frac{1}{n}\log\frac{P_{\hatX_j^n|Y_j^nS^j}(\hatx_j^n|y_j^n,s^j)}{Q_{\hatX_j^n|X^n\bY^kS^k\hat{\bX}^{j-1}}(\hatx_j^n|x^n,\by^k,s^k,\hat{\bx}^{j-1})}\geq -\eta,~\forall~j\in[2:k]\Big\}\label{def:calA4},\\
\calA_5
&:=\Big\{g:R_1\geq \frac{1}{n}\log\frac{P_{X^n|S_1}(x^n|s_1)}{P_X^n(x^n)}-\eta\Big\}\label{def:calA5},\\
\calA_6
&:=\Big\{g:R_j-\sum_{l\in[j-1]}R_l\geq \frac{1}{n}\log \frac{P_{X^n|S^j}(x^n|s^j)}{P_{X^n|S^{j-1}}(x^n|s^{j-1})}-\eta,~\forall~j\in[2:k]\Big\}\label{def:calA6},\\
\calA_7&:=
\Big\{g:D_j\geq d_j(x^n,\hatx_j^n)~\forall~j\in[k]\Big\}
=\Big\{g:D_j\geq \frac{1}{n}\sum_{i\in[n]}d_j(x_i,\hatx_{j,i})~\forall~j\in[k]\Big\}\label{def:calA7}.
\end{align}
Then we have the following non-asymptotic upper bound on the probability of non-excess-distortion.
\begin{lemma}
\label{fbln}
Given any $(n,M^k)$-code satisfying \eqref{conditions} and any distortion levels $D^k$, we have
\begin{align}
\rmP_\rmc^{(n)}(D^k)
\leq \Pr\Big\{\bigcap_{i\in[7]}\calA_i\Big\}+(2k+2)\exp(-n\eta).
\end{align}
\end{lemma}
The proof of Lemma \ref{fbln} is given in Appendix \ref{proof:fbln}.

In the remaining of this subsection, we single-letterize the bound in Lemma \ref{fbln}. Recall that given any $(i,j)\in[n]\times[k]$, we use $Y_{j,1}^{j,i}$ to denote $(Y_{j,1},\ldots,Y_{j,i})$.Recalling that the distributions starting with $P$ are all induced by the joint distribution $P_G$ in \eqref{def:joint} and using the choice of auxiliary random variables $(W_{1,i},\ldots,W_{k,i},V_i)$, we have
\begin{align}
\nn&P_{X^n\bY^{k\setminus 1}S^{k\setminus 1}|Y_1^nS_1}(x^n,\by^{k\setminus 1},s^{k\setminus 1}|y_1^n,s_1)\\*
&=\prod_{i\in[n]}P_{X_iY_{2,i}^{k,i}S^{k\setminus 1}|X^{i-1},Y_{2,1}^{2,i-1},\ldots,Y_{k,1}^{k,i-1},Y_1^n,S_1}(x_i,y_{2,i}^{k,i},s^{k\setminus 1}|x^{i-1},y_{2,1}^{2,i-1},\ldots,y_{k,1}^{k,i-1},y_1^n,s_1)\\
&=\prod_{i\in[n]}P_{X_iY_{2,i}^{k,i}S^{k\setminus 1}|X^{i-1},Y_{1,1}^{1,i-1},\ldots,Y_{k,1}^{i-1},Y_{1,i},S_1}(x_i,y_{2,i}^{k,i},s^{k\setminus 1}|x^{i-1},y_{1,1}^{1,i-1},\ldots,y_{k,1}^{k,i-1},y_{1,i},s_1)\label{usemc1}\\
&=\prod_{i\in[n]}P_{X_iY_{2,i}^{k,i}W_{2,i}^{k,i}|Y_{1,i}W_{1,i}}(x_i,y_{2,i}^{k,i},w_{2,i}^{k,i}|y_{1,i},w_{1,i})
\end{align}
\begin{align}
P_{\hatX_j^n|Y_j^nS^j}(\hatx_j^n|y_j^n,s_j)
&=\prod_{i\in[n]}P_{\hatX_{j,i}|Y_{j,1}^{j,i}S^j}(\hatx_{j,i}|y_{j,1}^{j,i},s^j)\\
&=\prod_{i\in[n]}P_{\hatX_{j,i}|X^{i-1},Y_{1,1}^{1,i-1},\ldots,Y_{k,1}^{k,i-1},Y_{j,i},S^j}(\hatx_{j,i}|x^{i-1},y_{1,1}^{1,i-1},\ldots,y_{k,1}^{k,i-1},y_{j,i},s^j)\label{usemc2}\\
&=\prod_{i\in[n]}P_{\hatX_{j,i}|Y_{j,i}W_{1,i}^{k,i}}(\hatx_{j,i}|y_{j,i},w_{1,i}^{k,i}),\\
P_{X^n|S_1}(x^n|s_1)
&=\prod_{i\in[n]}P_{X_i|X^{i-1}S_1}(x_i|x^{i-1},S_1)\\
&=\prod_{i\in[n]}P_{X_i|X^{i-1}Y_{1,1}^{1,i-1},\ldots,Y_{k,1}^{k,i-1}S_1}(x_i|x^{i-1}y_{1,1}^{1,i-1},\ldots,y_{k,1}^{k,i-1},S_1)\label{usemc2.5}\\
&=\prod_{i\in[n]}P_{X_i|W_{1,i}}(x_i|w_{1,i})\\
P_{X^n|S^{j-1}}(x^n|s^{j-1})
&=\prod_{i\in[n]}P_{X_i|X^{i-1}S^{j-1}}(x_i|x^{i-1},s^{j-1})\\
&=\prod_{i\in[n]}P_{X_i|X^{i-1}Y_{1,1}^{1,i-1},\ldots,Y_{k,1}^{k,i-1},S^{j-1}}(x_i|x^{i-1},y_{1,1}^{1,i-1},\ldots,y_{k,1}^{k,i-1},s^{j-1})\label{usemc3}\\
&=\prod_{i\in[n]}P_{X_i|W_{1,i}^{j-1,i}}(x_i,w_{1,i}^{j-1,i}),\\
P_{X^n|S^j}(x^n|s^j)
&=\prod_{i\in[n]}P_{X_i|W_{1,i}^{j,i}}(x_i,w_{1,i}^{j,i}),
\end{align}
where 
\eqref{usemc1} follows from the Markov chain $(X_i,Y_{2,i}^{k,i},S_2^k)-(X^{i-1},Y_{1,1}^{1,i-1},\ldots,Y_{k,1}^{k,i-1},Y_{1,i},S_1)-Y_{1,i+1}^{1,n}$, \eqref{usemc2} follows from the Markov chain $\hatX_{j,i}-(Y_{j,1}^{j,i},S^j)-(X^{i-1},Y_{1,1}^{1,i-1},\ldots,Y_{j-1,1}^{j-1,i-1},Y_{j+1,1}^{j+1,i-1},\ldots,Y_{k,1}^{k,i-1})$, \eqref{usemc2.5} follows from the Markov chain $X_i-(X^{i-1},S_1)-(Y_1^{i-1},\ldots,Y_k^{i-1})$, and \eqref{usemc3} follows from the Markov chain $X_i-(X^{i-1},S^{j-1})-(Y_{1,1}^{1,i-1},\ldots,Y_{k,1}^{k,i-1})$.

Furthermore, recall that for $i\in[n]$, $Q_{C_i|D_i}$ are arbitrary distributions where $C_i\in\calT_i$ and $D_i\in\calT_i$. Note that Lemma \ref{fbln} holds for arbitrary choices of distributions $Q_{C|D}$ where $C\in\calG$ and $D\in\calG$. The proof of Lemma \ref{fbl-sl} is completed by using Lemma \ref{fbln} with the following choices of auxiliary distributions and noting that $\calB_7=\calA_7$:
\begin{align}
Q_{X^n}(x^n)&:=\prod_{i\in[n]}Q_{X_i}(x_i),\\
Q_{\bY^k|X^nS^k}(\by^k|x^n,s^k)
&:=\prod_{i\in[n]}Q_{Y_{1,i}^{k,i}|X_i,W_{1,i}^{k,i}}(y_{1,i}^{k,i}|x_i,w_{1,i}^{k,i}),\\
Q_{X^n\bY^{k\setminus 1}S_2^k|Y_1^nS_1\hatX_1^n}(x^n,\by^{k\setminus 1},s_2^k|y_1^n,s_1,\hatx_1^n)
&:=\prod_{i\in[n]}Q_{X_iY_{2,i}^{k,i}W_{2,i}^{k,i}|Y_{1,i}W_{1,i}\hatX_{1,i}}
(x_i,y_{2,i}^{k,i},w_{2,i}^{k,i}|y_{1,i},w_{1,i},\hatx_{1,i})\\
Q_{\hatX_j^n|X^n\bY^kS^k\hat{\bX}^{j-1}}(\hatx_j^n|x^n,\by^k,s^k,\hat{\bx}^{k\setminus j})
&:=\prod_{i\in[n]}Q_{\hatX_{j,i}|X_i,Y_{1,i}^{k,i},W_{1,i}^{k,i},\hatX_{1,i}^{j-1,i}}(\hatx_{j,i}|x_i,y_{1,i}^{k,i},w_{1,i}^{k,i},\hatx_{1,i}^{j-1,i}).
\end{align}

\subsection{Proof of Lemma \ref{fbln}}
\label{proof:fbln}
Recall the definition of the probability of non-excess-distortion $\rmP_\rmc^{(n)}(D^k)$ in \eqref{def:pcn} and the definitions of sets $\{\calA_j\}_{j\in[7]}$ in \eqref{def:calA1} to \eqref{def:calA7}. For any $(n,M^k)$-code, we have that
\begin{align}
\rmP_\rmc^{(n)}(D^k)
&=\Pr\{\calA_7\}\\
&=\Pr\bigg\{\calA_7\bigcap(\bigcap_{j\in[6]}\calA_j)\bigg\}+\Pr\bigg\{\calA_7\bigcap(\bigcup_{j\in[6]}\calA_j^\rmc)\bigg\}\\
&=\Pr\Big\{\bigcap_{j\in[7]}\calA_j\Big\}+\sum_{j\in[6]}\Pr\{\calA_j^\rmc\}\label{useub},
\end{align}
where \eqref{useub} follows from the union bound and the fact that $\Pr\{\calA\cap\calB\}\leq \Pr\{\calB\}$ for any two sets $\calA$ and $\calB$. The proof of Lemma \ref{fbln} is completed by showing that 
\begin{align}
\sum_{j\in[6]}\Pr\{\calA_j^\rmc\}\leq (2k+2)\exp(-n\eta)\label{toshow}.
\end{align}
In the remaining of this subsection, we show that \eqref{toshow} holds. Recall the joint distribution of $G$ in \eqref{def:joint}. In the following, when we use a (conditional) distribution starting with $P$, we mean that the (conditional) distribution is induced by the joint distribution $P_G$ in \eqref{def:joint}.

Using the definition of $\calA_1$ in \eqref{def:calA1}, we have
\begin{align}
\Pr\{\calA_1^\rmc\}
&=\sum_{x^n\in\calX^n} P_X^n(x^n)1\{P_X^n(x^n)\leq \exp(-n\eta)Q_{X^n}(x^n)\}\\
&\leq \exp(-n\eta)\label{showa1}.
\end{align}
Similarly to \eqref{showa1}, we have that
\begin{align}
\Pr\{\calA_2^\rmc\}
&=\sum_{g\in\calA_2^\rmc}P_G(g)\\
&=\sum_{x^n,s^k,\by^k}P_{XY^k}(x^n,\by^k)\big(\prod_{j\in[k]}P_{S_j|X^n}(s_j|x^n)\big)
1\{P_{Y^k|X}^n(\by^k|x^n)\leq \exp(-n\eta)Q_{\bY^k|X^nS^k}(\by^k|x^n,s^k)\}\\
&\leq \exp(-n\eta)\sum_{x^n,s^k,\by^k}P_X^n(x^n)Q_{\bY^k|X^nS^k}(\by^k|x^n,s^k)\big(\prod_{j\in[k]}P_{S_j|X^n}(s_j|x^n)\big)\\
&\leq \exp(-n\eta),\\
\Pr\{\calA_3^\rmc\}
&=\sum_{g\in\calA_3^\rmc}P_G(g)\\
&\leq\exp(-n\eta)\sum_{x^n,\by^k,s^k,\hatx_1^n}P_{Y_1^nS_1}(y_1^n,s_1)P_{\hatX_1^n|Y_1^nS_1}(\hatx_1^n|y_1^n,s_1)Q_{X^n\bY^{k\setminus 1}S_2^k|Y_1^nS_1\hatX_1^n}(x^n,\by^{k\setminus 1},s_2^k|y_1^n,s_1,\hatx_1^n)\\
&\leq \exp(-n\eta),
\end{align}

Furthermore, using the definition of $\calA_4$ in \eqref{def:calA4} and the union bound, we have that
\begin{align}
\Pr\{\calA_4^\rmc\}
&\leq \sum_{j\in[2:k]}\exp(-n\eta)\sum_{x^n,\by^k,s^k,\hat{\bx}^j}P_{XY^k}^n(x^n,\by^k)\big(\prod_{l\in[k]}P_{S_l|X^n}(s_l|x^n)\big)\big(\prod_{l\in[j-1]}P_{\hatX_l^n|Y_l^nS^l}(\hatx_l^n|y_l^n,s_l)\big)\\*
&\qquad\qquad\times Q_{\hatX_j^n|X^n\bY^kS^k\hat{\bX}^{j-1}}(\hatx_j^n|x^n,\by^k,s^k,\hat{\bx}^{j-1})\\
&\leq (k-1)\exp(-n\eta)\label{unioncala4c}.
\end{align}

Furthermore, using the definition of $\calA_5$ in \eqref{def:calA5}, we obtain that
\begin{align}
\Pr\{\calA_5^\rmc\}
&\leq\sum_{x^n,s_1}P_{S_1|X^n}(s_1|x^n)\exp(-n(R_1+\eta))P_{X^n|S_1}(x^n|s_1)\\
&\leq \sum_{x^n,s_1}\exp(-n(R_1+\eta))P_{X^n|S_1}(x^n|s_1)\label{ple1}\\
&=\sum_{s_1}\exp(-n(\eta+R_1))\\
&\leq\exp(-n\eta)\label{sizes_1},
\end{align}
where \eqref{ple1} follows since $P_{S_1|X^n}(s_1|x^n)\leq 1$ for all $(x^n,s_1)$, and \eqref{sizes_1} follows since $\sum_{s_1}=|\calW_1|=M_1\leq \exp(nR_1)$.

Using the definition of $\calA_6$ in \eqref{def:calA6} and the union bound similarly to \eqref{unioncala4c}, we have
\begin{align}
\Pr\{\calA_6^\rmc\}
&\leq\sum_{j\in[2:k]}\sum_{x^n,s^j}P_{S^{j-1}}(s^{j-1})\exp(-n\eta)P_{X^n|S^j}(x^n|s^j)\exp(-n(R_j-\sum_{l\in[j-1]}R_l))P_{S_j|X^n}(s_j|x^n)\\
&\leq \sum_{j\in[2:k]}\exp(-n\eta)\sum_{x^n,s^j}P_{S^{j-1}}(s^{j-1})P_{X^n|S^j}(x^n|s^j)\exp(-n(R_j-\sum_{l\in[j-1]}R_l))\label{ps2le1}\\
&\leq \sum_{j\in[2:k]}\exp(-n\eta)\sum_{s_j}\exp(-n(R_j-\sum_{l\in[j-1]}R_l))\\
&\leq(k-1)\exp(-n\eta)\label{sizes_2},
\end{align}
where \eqref{ps2le1} follows since $P_{S_j|X^n}(s_j|x^n)\leq 1$ for all $(x^n,s_j)$ and \eqref{sizes_2} follows since $\sum_{s_j}=|\calM_j|=M_j\leq \exp(n(R_j-\sum_{l\in[j-1]}R_l))$.

\subsection{Proof of Lemma \ref{fbletype}}
\label{proof:fbletype}

For any $(\mu,\alpha^k,\beta^k)\in\bbR_+\times [0,1]^{2k}$ satisfying \eqref{linearconstraints}, for $i\in[4]$, define $\calF_i=\calB_i$ (cf. \eqref{def:calB1} to \eqref{def:calB4}) and for $i\in[5:7]$, define
\begin{align}
\calF_5&:=\bigg\{
g:\mu\alpha_1R_1\geq \frac{\mu\alpha_1}{n}\sum_{i\in[n]}\log\frac{Q_{X_i|W_{1,i}}(x_i|w_{1,i})}{P_X(x_i)}-\mu\alpha_1\eta\bigg\}\label{def:calF5},\\
\calF_6&:=\bigg\{
g:\mu\alpha_j(R_j-\sum_{l\in[j-1]}R_l)\geq \sum_{i\in[n]}\frac{\mu\alpha_j}{n}\log\frac{Q_{X_i|W_{1,i}W_{2,i}}(x_i|w_{1,i},w_{2,i})}{P_{X_i|W_{1,i}}(x_i|w_{1,i})}-\mu\alpha_j\eta,~\forall~j\in[2:k]\bigg\}\label{def:calF6},\\
\calF_7&:=\bigg\{
g:\mu\beta_jD_j\geq \frac{\mu\beta_j}{n}\sum_{i\in[n]}\log\exp(d_1(x_i,\hatx_{1,i})),~\forall j\in[k]\bigg\}\label{def:calF8}.
\end{align}
Furthermore, let
\begin{align}
c(\mu,\alpha^k)&:=k+2+\sum_{j\in[k]}\mu\alpha_j\label{def:cmab}.
\end{align}

Using Lemma \ref{fbl-sl} and definitions in \eqref{def:calF5} to \eqref{def:cmab}, we obtain that
\begin{align}
\nn&\rmP_\rmc^{(n)}(D^k)-(2k+2)\exp(-n\eta)\\*
&\leq \Pr\Big\{\bigcap_{i\in[7]}\calF_i\Big\}\\
&\leq \Pr\Big\{n(\mu\kappa^{(\alpha^k,\beta^k)}(R^k,D^k)+c(\mu,\alpha^k)\eta)\geq \sum_{i\in[n]}\log f_{Q_i,P_i}^{(\mu,\alpha^k,\beta^k)}(T_i)\Big\}\\
&\leq \exp\bigg\{n\lambda(\mu\kappa^{(\alpha^k,\beta^k)}(R^k,D^k)+c(\mu,\alpha^k)\eta)+\log\mathbb{E}\Big[\exp\Big(-\lambda \sum_{i\in[n]}\log f_{Q_i,P_i}^{(\mu,\alpha^k,\beta^k)}(T_i)\Big)\Big]\bigg\}\label{usecramer}\\
&=\exp\bigg\{n\Big(\lambda\mu\kappa^{(\alpha^k,\beta^k)}(R^k,D^k)+\lambda c(\mu,\alpha^k)\eta-\frac{1}{n}\Omega^{(\lambda,\mu,\alpha^k,\beta^k)}(\{P_i,Q_i\}_{i\in[n]})\Big)\bigg\}\label{usedefinitions2},
\end{align}
where \eqref{usecramer} follows from Cram\'er's bound in \cite[Lemma 13]{zhou2016cilossy} and \eqref{usedefinitions2} follows from the definition of $\Omega^{(\lambda,\mu,\alpha^k,\beta^k)}(\{P_i,Q_i\}_{i\in[n]})$ in \eqref{def:OmegaQiPi}.

Choose $\eta$ such that
\begin{align}
-\eta=\lambda\mu\kappa^{(\alpha^k,\beta^k)}(R^k,D^k)+\lambda c(\mu,\alpha^k)\eta-\frac{1}{n}\Omega^{(\lambda,\mu,\alpha^k,\beta^k)}(\{P_i,Q_i\}_{i\in[n]}),
\end{align}
i.e.,
\begin{align}
\eta=\frac{\frac{1}{n}\Omega^{(\lambda,\mu,\alpha^k,\beta^k)}(\{P_i,Q_i\}_{i\in[n]})-\lambda \mu\kappa^{(\alpha^k,\beta^k)}(R^k,D^k)}{1+\lambda c(\mu,\alpha^k)}\label{chooseeta}.
\end{align}
The proof of Lemma \ref{fbletype} is completed by combining \eqref{usedefinitions2} and \eqref{chooseeta}.

\subsection{Proof of Lemma \ref{vital}}
\label{proof:vital}
Recall that for each $i\in[n]$, we use $t_i$ to denote $(x_i,y_{1,i}^{k,i},w_{1,i}^{k,i},\hatx_{1,i}^{k,i})$ and use $T_i$ similarly.

Recall that the auxiliary random variables are chosen as $w_{1,i}=(x^{i-1},y_1^{i-1},\ldots,y_k^{i-1},s_1)$ and $w_{j,i}=s_j$ for all $j\in[2:k]$. Using the definition of $f_{Q_i,P_i}^{(\mu,\alpha^k,\beta^k)}$ in \eqref{def:fQPimabg}, we define
\begin{align}
h_{Q_i,P_i}^{(\lambda,\mu,\alpha^k,\beta^k)}(t_i)&:=\exp\Big(-\lambda \log f_{Q_i,P_i}^{(\mu,\alpha^k,\beta^k)}(t_i)\Big)\label{def:hQiPi}.
\end{align}
Recall the joint distribution of $G$ in \eqref{def:joint}. For each $j\in[n]$, define
\begin{align}
\tilC_j
&:=\sum_{g}P_G(g)\prod_{i\in[j]}h_{Q_i,P_i}^{(\lambda,\mu,\alpha^k,\beta^k)}(t_i),\label{def:tilCi}\\
P_G^{(\lambda,\mu,\alpha^k,\beta^k)|j}(g)
&:=\frac{P_G(g)\prod_{i\in[j]}h_{Q_i,P_i}^{(\lambda,\mu,\alpha^k,\beta^k)}(t_i)}{\tilC_i}\label{def:PGi},\\
\Lambda_j^{(\lambda,\mu,\alpha^k,\beta^k)}(\{Q_i,P_i\}_{i\in[n]})
&:=\frac{\tilC_j}{\tilC_{j-1}}\label{def:Lambdai}.
\end{align}
Combining \eqref{def:OmegaQiPi} and \eqref{def:Lambdai}, we have
\begin{align}
\exp\Big(-\Omega^{(\lambda,\mu,\alpha^k,\beta^k)}_{(\{P_i,Q_i\}_{i\in[n]})}\Big)
&=\mathbb{E}\Big[\prod_{i\in[n]}h_{Q_i,P_i}^{(\lambda,\mu,\alpha^k,\beta^k)}(T_i)\Big]\\
&=\sum_{g\in\calG}P_G(g)\prod_{i\in[n]}h_{Q_i,P_i}^{(\lambda,\mu,\alpha^k,\beta^k)}(t_i)\\
&=\prod_{i\in[n]}\Lambda_i^{(\lambda,\mu,\alpha^k,\beta^k)}(\{Q_i,P_i\})\label{expOmega=}.
\end{align}

Furthermore, similar to \cite[Lemma 5]{oohama2018exponential}, we obtain the following lemma, which is critical in the proof of Lemma \ref{vital}.
\begin{lemma}
\label{lemma:relate}
For each $j\in[n]$,
\begin{align}
\Lambda_j^{(\lambda,\mu,\alpha^k,\beta^k)}(\{Q_i,P_i\}_{i\in[n]})
&=\sum_{g\in\calG}P_G^{(\lambda,\mu,\alpha^k,\beta^k)|j-1}(g)h_{Q_j,P_j}^{(\mu,\alpha^k,\beta^k)}(t_i)\label{relateLambda}.
\end{align}
\end{lemma}
Furthermore, for each $j\in[n]$, define
\begin{align}
P^{(\lambda,\mu,\alpha^k,\beta^k)}(t_j)
&:=\sum_{\substack{x_{j+1}^n,y_{1,j+1}^n,\ldots,y_{k,j+1}^n,\\\hatx_1^{j-1},\ldots,\hatx_k^{j-1},\hatx_{1,j+1}^n,\dots,\hatx_{k,j+1}^n}}P_G^{(\lambda,\mu,\alpha^k,\beta^k)|j-1}(g)\label{def:pi}.
\end{align}
Using Lemma \ref{lemma:relate} and \eqref{def:pi}, we have that for each $j\in[n]$,
\begin{align}
\Lambda_j^{(\lambda,\mu,\alpha^k,\beta^k)}(\{Q_i,P_i\}_{i\in[n]})
&=\sum_{t_j}P^{(\lambda,\mu,\alpha^k,\beta^k)}(t_j)h_{Q_j,P_j}^{(\mu,\alpha^k,\beta^k)}(t_j)\label{alter:Lambdaj}.
\end{align}

Recall that the auxiliary distributions $\{Q_i\}_{i\in[n]}$ can be arbitrary distributions. Following the recursive method in \cite{oohama2018exponential}, for each $i\in[n]$, we choose $Q_i$ such that
\begin{align}
Q_i(t_i)=P^{(\lambda,\mu,\alpha^k,\beta^k)}(t_i).
\end{align}
Let $Q_{C_i|D_i}$, where $C_i\in\calT_i$ and $D_i\in\calT_i$, be induced by $Q_{i}$.
Using the definition of $h_{Q_i,P_i}^{(\lambda,\mu,\alpha^k,\beta^k)}(t_i)$ in \eqref{def:hQiPi}, we define
\begin{align}
\xi_{Q_i,P_i}^{(\lambda,\mu,\alpha^k,\beta^k)}(t_i)
\nn&:=h_{Q_i,P_i}^{(\lambda,\mu,\alpha^k,\beta^k)}(t_i)
\Bigg(\frac{P_{X_iY_{2,i}^{k,i}W_{2,i}^{k,i}|Y_{1,i}W_{1,i}}(x_i,y_{2,i}^{k,i},w_{2,i}^{k,i}|y_{1,i},w_{1,i})}{Q_{X_iY_{2,i}^{k,i}W_{2,i}^{k,i}|Y_{1,i}W_{1,i}}(x_i,y_{2,i}^{k,i},w_{2,i}^{k,i}|y_{1,i},w_{1,i})}\Bigg)^{-\lambda}\\*
\nn&\qquad\times \Bigg(\prod_{j\in[2:k]}\frac{P_{\hatX_{j,i}|Y_{j,i}W_{1,i}^{j,i}}(\hatx_{j,i}|y_{j,i},w_{1,i}^{j,i})}{Q_{\hatX_{j,i}|Y_{j,i}W_{1,i}^{j,i}}(\hatx_{j,i}|y_{j,i},w_{1,i}^{j,i})}\Bigg)^{-\lambda} \Bigg(\frac{P_{X_i|W_{1,i}}(x_i|w_{1,i})}{Q_{X_i|W_{1,i}}(x_i|w_{1,i})}\Bigg)^{-\lambda\mu\alpha_1}\\*
&\qquad\times \prod_{j\in[2:k]}\Bigg(\frac{P_{X_i|W_{1,i}^{j-1,i}}(x_i,w_{1,i}^{j-1,i})}{Q_{X_i|W_{1,i}^{j-1,i}}(x_i,w_{1,i}^{j-1,i})}\Bigg)^{-\lambda\mu\alpha_j}
\label{def:lQiPi}.
\end{align}
In the following, for simplicity, we let $\Psi:=1-k\lambda-\sum_{j\in[k]}\lambda\mu\alpha_j$. Combining \eqref{def:pi} and \eqref{alter:Lambdaj}, we obtain that for each $l\in[n]$,
\begin{align}
\nn&\Lambda_l^{(\lambda,\mu,\alpha^k,\beta^k)}(\{Q_i,P_i\}_{i\in[n]})\\*
&=\mathbb{E}_{Q_l}\big[h_{Q_l,P_l}^{(\mu,\alpha^k,\beta^k)}(T_l)\big]\\
\nn&=\mathbb{E}_{Q_l}\Bigg[\xi_{Q_l,P_l}^{(\mu,\alpha^k,\beta^k)}(T_l)
\Bigg(\frac{P_{X_lY_{2,l}^{k,l}W_{2,l}^{k,l}|Y_{1,l}W_{1,l}}(x_l,y_{2,l}^{k,l},w_{2,l}^{k,l}|y_{1,l},w_{1,l})}{Q_{X_lY_{2,l}^{k,l}W_{2,l}^{k,l}|Y_{1,l}W_{1,l}}(x_l,y_{2,l}^{k,l},w_{2,l}^{k,l}|y_{1,l},w_{1,l})}\Bigg)^{\lambda}\Bigg(\prod_{j\in[2:k]}\frac{P_{\hatX_{j,l}|Y_{j,l}W_{1,l}^{j,l}}(\hatx_{j,l}|y_{j,l},w_{1,l}^{j,l})}{Q_{\hatX_{j,l}|Y_{j,l}W_{1,l}^{j,l}}(\hatx_{j,l}|y_{j,l},w_{1,l}^{j,l})}\Bigg)^{\lambda}\\*
&\qquad\qquad\times \Bigg(\frac{P_{X_l|W_{1,l}}(x_l|w_{1,l})}{Q_{X_l|W_{1,l}}(x_l|w_{1,l})}\Bigg)^{\lambda\mu\alpha_1}\prod_{j\in[2:k]}\Bigg(\frac{P_{X_l|W_{1,l}^{j-1,l}}(x_l,w_{1,l}^{j-1,l})}{Q_{X_l|W_{1,l}^{j-1,l}}(x_l,w_{1,l}^{j-1,l})}\Bigg)^{\lambda\mu\alpha_j}\Bigg]\\
&\leq \bigg(\mathbb{E}_{Q_l}\Big[\Big(\xi_{Q_l,P_l}^{(\mu,\alpha^k,\beta^k)}(T_l)\Big]\Big)^{\frac{1}{\Psi}}\bigg)^{\Psi}
\Bigg(\mathbb{E}\Bigg[\frac{P_{X_lY_{2,l}^{k,l}W_{2,l}^{k,l}}|Y_{1,l}W_{1,l}(x_l,y_{2,l}^{k,l},w_{2,l}^{k,l}|y_{1,l},w_{1,l})}{Q_{X_lY_{2,l}^{k,l}W_{2,l}^{k,l}}|Y_{1,l}W_{1,l}(x_l,y_{2,l}^{k,l},w_{2,l}^{k,l}|y_{1,l},w_{1,l})}\Bigg]\Bigg)^{\lambda}
\\*
\nn&\qquad\times\prod_{j\in[2:k]}\Bigg(\mathbb{E}\Bigg[\frac{P_{\hatX_{j,l}|Y_{j,l}W_{1,l}^{j,l}}(\hatx_{j,l}|y_{j,l},w_{1,l}^{j,l})}{Q_{\hatX_{j,l}|Y_{j,l}W_{1,l}^{j,l}}(\hatx_{j,l}|y_{j,l},w_{1,l}^{j,l})}\Bigg]\Bigg)^{\lambda}\Bigg(\mathbb{E}\Bigg[\frac{P_{X_l|W_{1,l}}(x_l|w_{1,l})}{Q_{X_l|W_{1,l}}(x_l|w_{1,l})}\Bigg]\Bigg)^{\lambda\mu\alpha_1}\\*
&\qquad\times\prod_{j\in[2:k]}\Bigg(\mathbb{E}\Bigg[\frac{P_{X_l|W_{1,l}^{j-1,l}}(x_l,w_{1,l}^{j-1,l})}{Q_{X_l|W_{1,l}^{j-1,l}}(x_l,w_{1,l}^{j-1,l})}\Bigg]\Bigg)^{\lambda\mu\alpha_j}
\label{useholder4Psi}\\
&\leq \exp\bigg(-\Psi\Omega^{(\frac{\lambda}{\Psi},\mu,\alpha^k,\beta^k)}(Q_j)\bigg)\label{usedefOmegaQ}\\
&=\exp\bigg(-\frac{\Omega^{(\theta,\mu,\alpha^k,\beta^k)}(Q_j)}{1+k\theta+\sum_{j\in[k]}\theta\mu\alpha_j}\bigg)\label{usethetalambda}\\
&\leq\exp\bigg(-\min_{Q_j\in\calP(\calT_j)}\frac{\Omega^{(\theta,\mu,\alpha^k,\beta^k)}(Q_j)}{1+k\theta+\sum_{j\in[k]}\theta\mu\alpha_j}\bigg)\label{optimization}\\
&=\exp\bigg(-\frac{\Omega^{(\theta,\mu,\alpha^k,\beta^k)}}{1+k\theta+\sum_{j\in[k]}\theta\mu\alpha_j}\bigg)\label{cardinalityandusedefinitions}
\end{align}
where \eqref{useholder4Psi} follows H\"older's inequality, \eqref{usedefOmegaQ} follows from the definitions of $\Omega^{(\theta,\mu,\alpha^k,\beta^k)}(\cdot)$ in \eqref{def:OmegaQT} and $\xi_{Q_j,P_j}^{(\mu,\alpha^k,\beta^k)}(\cdot)$ in \eqref{def:lQiPi}, \eqref{usethetalambda} follows from the result in \eqref{def:theta} and \eqref{lambda=f(theta)}, and \eqref{cardinalityandusedefinitions} follows from the definition of $\Omega^{(\theta,\mu,\alpha^k,\beta^k)}$ in \eqref{def:Omega} and the fact it is sufficient to consider distributions $Q_j$ with cardinality bounds $W_{1,j}\leq |\calX|$ and $W_{2,j}\leq |\calX|^2$ for the optimization problem in \eqref{optimization} (the proof of this fact is similar to \cite[Property 4(a)]{oohama2018exponential} and thus omitted).

The proof of Lemma \ref{vital} is completed by combining \eqref{expOmega=} and \eqref{cardinalityandusedefinitions}.

\subsection{Proof of Lemma \ref{relateF&tilF}}
\label{proof:relateF&tilF}

\subsubsection{Proof of Claim (i)}
For any $Q_T\in\calQ$ (see \eqref{def:calQ}), let $P_T\in\calP_{\rm{sh}}$ (see \eqref{def:calpsh}) be chosen such that $P_{W^k|X}=Q_{W^k|X}$ and $P_{\hatX_j|Y_jW^j}=Q_{\hatX_j|Y_jW^j}$ for all $j\in[k]$.

In the following, we drop the subscript of distributions when there is no confusion. For any $(\theta,\mu,\alpha^k,\beta^k)\in\bbR_+^2\times[0,1]^{2k}$ satisfying \eqref{linearconstraints} and
\begin{align}
\sum_{j\in[2:k]}\mu\alpha_j\leq 1~\mathrm{and~}\forall~l\in[k],~\theta\leq \frac{1}{1+\mu\alpha_l},\label{constraints:parameters}
\end{align}
using the definition of $\Omega^{(\theta,\mu,\alpha^k,\beta^k)}(Q_T)$ in \eqref{def:OmegaQT}, we obtain 
\begin{align}
\nn&\exp\Big(-\Omega^{(\theta,\mu,\alpha^k,\beta^k)}(Q_T)\Big)\\*
\nn&=\mathbb{E}_{Q_T}
\Bigg[\bigg(\frac{P(X,Y^k)Q(X,Y^{k\setminus 1},W^{k\setminus 1}|Y_1,W_1)\big(\prod_{j\in[2:k]}Q(\hatX_j|Y,W^j)\big)}{Q(X)Q(Y^k|X,W^k)Q(X,Y^{k\setminus 1},W^{k\setminus 1}|Y_1,W_1,\hatX_1)\big(\prod_{j\in[2:k]}Q(\hatX_j|X,Y^k,W^k,\hatX^{j-1})\big)}\bigg)^\theta\\*
&\qquad\qquad\times \bigg(\frac{P(X)}{Q(X|W_1)}\bigg)^{\theta\mu\alpha_1}
\bigg(\prod_{j\in[2:k]}\bigg(\frac{Q(X|W^{j-1})}{Q(X|W^j)}\bigg)^{\theta\mu\alpha_j}\bigg)
\exp\Big(-\theta\mu\big(\sum_{j\in[k]}\beta_j d_j(X,\hatX_j)\big)\Big)\Bigg]\\
&=\mathbb{E}_{Q_T}\Bigg[\bigg(\frac{P(T)}{Q(T)}\bigg)^{\theta}\bigg(\frac{P(X)}{Q(X|W_1)}\bigg)^{\theta\mu\alpha_1}
\bigg(\prod_{j\in[2:k]}\bigg(\frac{Q(X|W^{j-1})}{Q(X|W^j)}\bigg)^{\theta\mu\alpha_j}\bigg)
\exp\Big(-\theta\mu\big(\sum_{j\in[k]}\beta_j d_j(X,\hatX_j)\big)\Big)\Bigg]\label{whyfollow}\\
&=\mathbb{E}_{Q_T}\Bigg[\bigg(\frac{P(T)}{Q(T)}\bigg)^{\theta}\bigg(\frac{P(X)}{P(X|W_1)}\bigg)^{\theta\mu\alpha_1}
\bigg(\prod_{j\in[2:k]}\bigg(\frac{Q(X|W^{j-1})}{P(X|W^j)}\bigg)^{\theta\mu\alpha_j}\bigg)
\exp\Big(-\theta\mu\big(\sum_{j\in[k]}\beta_j d_j(X,\hatX_j)\big)\Big)\Bigg]\\*
&\qquad\qquad\times\bigg(\prod_{j\in[k]}\bigg(\frac{P(X|W^j)}{Q(X|W^j)}\bigg)^{\theta\mu\alpha_j}\bigg)\\
\nn&\leq \Bigg(\mathbb{E}_{Q_T}\Bigg[\bigg(\frac{P(T)}{Q(T)}\bigg)\bigg(\frac{P(X)}{P(X|W_1)}\bigg)^{\mu\alpha_1}
\bigg(\prod_{j\in[2:k]}\bigg(\frac{Q(X|W^{j-1})}{P(X|W^j)}\bigg)^{\mu\alpha_j}\bigg)
\exp\Big(-\mu\big(\sum_{j\in[k]}\beta_j d_j(X,\hatX_j)\big)\Big)\Bigg]\Bigg)^\theta\\*
&\qquad\qquad\times\prod_{j\in[k]}\bigg(\mathbb{E}_{Q_T}\Bigg[\bigg(\frac{P(X|W^j)}{Q(X|W^j)}\bigg)^{\frac{\theta\mu\alpha_j}{1-\theta}}\bigg]\bigg)^{1-\theta}\label{useHineq}\\
&\leq \Bigg(\mathbb{E}_{P_T}\Bigg[\bigg(\frac{P(X)}{P(X|W_1)}\bigg)^{\mu\alpha_1}
\bigg(\prod_{j\in[2:k]}\bigg(\frac{Q(X|W^{j-1})}{P(X|W^j)}\bigg)^{\mu\alpha_j}\bigg)
\exp\Big(-\mu\big(\sum_{j\in[k]}\beta_j d_j(X,\hatX_j)\big)\Big)\Bigg]\Bigg)^\theta\label{useconcave}\\
\nn&=\Bigg(\mathbb{E}_{P_T}\Bigg[\bigg(\frac{P(X)}{P(X|W_1)}\bigg)^{\mu\alpha_1}
\bigg(\prod_{j\in[2:k]}\bigg(\frac{P(X|W^{j-1})}{P(X|W^j)}\bigg)^{\mu\alpha_j}\bigg)
\exp\Big(-\mu\big(\sum_{j\in[k]}\beta_j d_j(X,\hatX_j)\big)\Big)\\*
&\qquad\qquad\qquad\times \prod_{j\in[2:k]}\bigg(\frac{Q(X|W^{j-1})}{P(X|W^{j-1})}\bigg)^{\mu\alpha_j}\Bigg]\Bigg)^\theta\\
\nn&=\Bigg(\mathbb{E}_{P_T}\Bigg[\bigg(\bigg(\frac{P(X)}{P(X|W_1)}\bigg)^{\mu\alpha_1}
\bigg(\prod_{j\in[2:k]}\bigg(\frac{P(X|W^{j-1})}{P(X|W^j)}\bigg)^{\mu\alpha_j}\bigg)
\exp\Big(-\mu\big(\sum_{j\in[k]}\beta_j d_j(X,\hatX_j)\big)\Big)\bigg)^{\frac{1}{1-\sum_{j\in[2:k]}\mu\alpha_j}}\Bigg]\Bigg)^{\theta(1-\sum_{j\in[2:k]}\mu\alpha_j)}\\*
&\qquad\qquad\qquad\times\prod_{j\in[2:k]}\Bigg(\mathbb{E}_{P_T}\Bigg[\bigg(\frac{Q(X|W^{j-1})}{P(X|W^{j-1})}\bigg)\Bigg]\Bigg)^{\theta\mu\alpha_j}\label{useholder2}\\
&=\exp\bigg(-\theta(1-\sum_{j\in[2:k]}\mu\alpha_j)\tilde{\Omega}\Big(\frac{\mu}{1-\sum_{j\in[2:k]}\mu\alpha_j},\alpha^k,\beta^k\Big)\bigg)\label{usedeftilOm},
\end{align}
where \eqref{whyfollow} follows since i) with our choice of $P_T\in\calP_{\rm{sh}}$, we have 
\begin{align}
P(T)=P(X,Y^k)P(W^k|X)\big(\prod_{j\in[k]}P(\hatX_j|Y_j,W^j)\big)
\end{align}
and ii) the following equality holds
\begin{align}
\frac{Q(X,Y^{k\setminus 1},W^{k\setminus 1}|Y_1,W_1)}{Q(X,Y^{k\setminus 1},W^{k\setminus 1}|Y_1,W_1,\hatX_1)}
&=\frac{Q(\hatX_1|Y_1,W_1)}{Q(\hatX_1|X,Y^k,W^k)},
\end{align}
\eqref{useHineq} follows from H\"older's inequality, \eqref{useconcave} follows from the concavity of $X^a$ for $a\in[0,1]$ and the choice of $\theta$ which ensures $\frac{\theta\mu\alpha_j}{1-\theta}\leq 1$ for all $j\in[k]$, \eqref{useholder2} follows by applying H\"older's inequality and recalling that $\sum_{j\in[2:k]}\mu\alpha_j\leq 1$, and \eqref{usedeftilOm} follows from the definition of $\tilde{\Omega}^{(\lambda,\alpha^k,\beta^k)}(P_T)$ in \eqref{def:tildeOmegaPT}.

Therefore, for any $(\theta,\mu,\alpha^k,\beta^k)\in\bbR_+^2\times[0,1]^{2k}$ satisfying \eqref{linearconstraints} and \eqref{constraints:parameters}, using the definition of $\Omega^{(\theta,\mu,\alpha^k,\beta^k)}$ in \eqref{def:Omega} and the result in \eqref{usedeftilOm}, we have that
\begin{align}
\Omega^{(\theta,\mu,\alpha^k,\beta^k)}\geq \theta(1-\sum_{j\in[2:k]}\mu\alpha_j)\tilde{\Omega}\Big(\frac{\mu}{1-\sum_{j\in[2:k]}\mu\alpha_j},\alpha^k,\beta^k\Big)\label{r1touse}.
\end{align}

Recalling the definition of $F(R^k,D^k)$ in \eqref{def:F} and using the result in \eqref{r1touse}, we have
\begin{align}
\nn&F(R^k,D^k)\\*
&=\sup_{(\theta,\mu,\alpha^k,\beta^k)\in\bbR_+^2\times[0,1]^{2k}:~\eqref{linearconstraints}}\frac{\Omega^{(\theta,\mu,\alpha^k,\beta^k)}-\theta\mu\kappa^{(\alpha^k,\beta^k)}(R^k,D^k)}{1+(2k+2)\theta+\sum_{j\in[k]}2\theta\mu\alpha_j}\\
&\geq \sup_{\substack{(\theta,\mu,\alpha^k,\beta^k)\in\bbR_+^2\times[0,1]^{2k}:\\\eqref{linearconstraints}\mathrm{~and~}\eqref{constraints:parameters}}}\frac{\theta(1-\sum_{j\in[2:k]}\mu\alpha_j)\tilde{\Omega}\Big(\frac{\mu}{1-\sum_{j\in[2:k]}\mu\alpha_j},\alpha^k,\beta^k\Big)-\theta\mu\kappa^{(\alpha^k,\beta^k)}(R^k,D^k)}{1+(2k+2)\theta+\sum_{j\in[k]}2\theta\mu\alpha_j}\\
&=\sup_{\substack{(\mu,\alpha^k,\beta^k)\in\bbR_+\times[0,1]^{2k}:\\\eqref{linearconstraints}~\mathrm{and~}\mu\leq\frac{1}{\sum_{j\in[2:k]}\alpha_j}}}\sup_{\theta\in\bbR_+:\max_{j\in[k]}\theta(1+\mu\alpha_j)\leq 1}\frac{\theta(1-\sum_{j\in[2:k]}\mu\alpha_j)\tilde{\Omega}\Big(\frac{\mu}{1-\sum_{j\in[2:k]}\mu\alpha_j},\alpha^k,\beta^k\Big)-\theta\mu\kappa^{(\alpha^k,\beta^k)}(R^k,D^k)}{1+(2k+2)\theta+\sum_{j\in[k]}2\theta\mu\alpha_j}\\
&=\sup_{\substack{(\mu,\alpha^k,\beta^k)\in\bbR_+\times[0,1]^{2k}:\\\eqref{linearconstraints}~\mathrm{and~}\mu\leq\frac{1}{\sum_{j\in[2:k]}\alpha_j}}}\frac{(1-\sum_{j\in[2:k]}\mu\alpha_j)\tilde{\Omega}\Big(\frac{\mu}{1-\sum_{j\in[2:k]}\mu\alpha_j},\alpha^k,\beta^k\Big)-\mu\kappa^{(\alpha^k,\beta^k)}(R^k,D^k)}{2k+3+\mu\alpha^+ +\sum_{l\in[k]}2\mu\alpha_l}\label{calculation}\\
&=\sup_{\substack{(\lambda,\alpha^k,\beta^k)\in\bbR_+\times[0,1]^{2k}:\\\eqref{linearconstraints}}}\frac{\tilde{\Omega}^{(\lambda,\alpha^k,\beta^k)}-\lambda\kappa^{(\alpha^k,\beta^k)}(R^k,D^k)}{2k+3+\lambda\alpha^+ +\sum_{j\in[2:k]}\lambda(2k+3)\alpha_j+\sum_{l\in[k]}2\lambda\alpha_l}\label{letlambda=}\\
&=\tilF(R^k,D^k)\label{usetilF},
\end{align}
where \eqref{calculation} follows since
\begin{align}
\sup_{\theta\in\bbR_+:\max_{j\in[k]}\theta(1+\mu\alpha_j)\leq 1}\frac{\theta}{1+(2k+2)\theta+\sum_{j\in[k]}2\theta\mu\alpha_j}
&=\min_{j\in [k]}\frac{1}{2k+3+\mu\alpha_j+\sum_{l\in[k]}2\mu\alpha_l}\\
&=\frac{1}{2k+3+\mu\alpha^+ +\sum_{l\in[k]}2\mu\alpha_l},
\end{align}
\eqref{letlambda=} follows by choosing $\lambda=\frac{\mu}{1-\sum_{j\in[2:k]}\mu\alpha_j}$ and \eqref{usetilF} follows from the definition of $\tilF$ in \eqref{def:tilF}.

\subsubsection{Proof of Claim (ii)}
Recall the definitions of $\tilde{\Omega}^{(\lambda,\alpha^k,\beta^k)}(P_T)$ in \eqref{def:tildeomegatPT} and $P_T^{(\lambda,\alpha^k,\beta^k)}$ in \eqref{def:tiltPT}. By simple calculation, one can verify that
\begin{align}
\frac{\partial \tilde{\Omega}^{(\lambda,\alpha^k,\beta^k)}(P_T)}{\partial \lambda}
&=\mathbb{E}_{P_T^{(\lambda,\alpha^k,\beta^k)}}\big[\tilde{\omega}^{(\alpha^k,\beta^k)}_{P_T}(T)\big],\label{firstd}\\
\frac{\partial^2 \tilde{\Omega}^{(\lambda,\alpha^k,\beta^k)}(P_T)}{\partial \lambda^2}
&=-\mathrm{Var}_{P_T^{(\lambda,\alpha^k,\beta^k)}}\big[\tilde{\omega}^{(\alpha^k,\beta^k)}_{P_T}(T)\big]\label{secondd}.
\end{align}

Applying Taylor expansion to $\tilde{\Omega}^{(\lambda,\alpha^k,\beta^k)}(P_T)$ at around $\lambda=0$ and combining \eqref{firstd}, \eqref{secondd}, we have that for any $P_T\in\calP_{\rm{sh}}$ and any $\lambda\in[1,\frac{1}{\sum_{j\in[k]}\alpha_j}]$, there exists $\tau\in[0,\lambda]$ such that
\begin{align}
\tilde{\Omega}^{(\lambda,\alpha^k,\beta^k)}(P_T)
&=\tilde{\Omega}^{(0,\alpha^k,\beta^k)}(P_T)+\lambda \mathbb{E}_{P_T^{(0,\alpha^k,\beta^k)}}\big[\tilde{\omega}^{(\alpha^k,\beta^k)}_{P_T}(T)\big]-\frac{\lambda^2}{2}\mathrm{Var}_{P_T^{(\tau,\alpha^k,\beta^k)}}\big[\tilde{\omega}^{(\alpha^k,\beta^k)}_{P_T}(T)\big]\\
&\geq \lambda \mathbb{E}_{P_T}\big[\tilde{\omega}^{(\alpha^k,\beta^k)}_{P_T}(T)\big]-\frac{\lambda^2\rho }{2}\label{usedefinitions},
\end{align}
where \eqref{usedefinitions} follows from the definitions in \eqref{def:tildeomegatPT}, \eqref{def:tiltPT} and \eqref{def:rho}.

Using the definitions in \eqref{def:tildeomegatPT}, \eqref{def:tildeOmega}, \eqref{def:Rabg} and the result in \eqref{usedefinitions}, we have that for any $\lambda\in[0,\frac{1}{\sum_{j\in[k]}\alpha_j}]$,
\begin{align}
\tilde{\Omega}^{(\lambda,\alpha^k,\beta^k)}
&=\min_{P_T\in\calP_{\rm{sh}}}\tilde{\Omega}^{(\lambda,\alpha^k,\beta^k)}(P_T)\\
&\geq \lambda\rmR^{(\alpha^k,\beta^k)}-\frac{\lambda^2\rho}{2}\label{result2touse}.
\end{align}

For any rate-distortion tuple outside the rate-distortion region, i.e., $(R^k,D^k)\notin\calR$, from Lemma \ref{rdregion:alternative}, we conclude that there exists $(\alpha^{k,*},\beta^{k,*})\in[0,1]^{2k}$ satisfying \eqref{linearconstraints} such that for some positive $\delta\in[0,\rho]$
\begin{align}
\kappa^{(\alpha^{k,^*},\beta^{k,*})}(R^k,D^k)
&\leq R^{(\alpha^{k,^*},\beta^{k,*})}-\delta\label{result3touse}.
\end{align}
Using the definition of $\tilF(R^k,D^k)$ in \eqref{def:tilF}, we have
\begin{align}
\tilF(R^k,D^k)
&=\sup_{\substack{(\lambda,\alpha^k,\beta^k)\in\bbR_+\times [0,1]^{2k}:~\eqref{linearconstraints}}}\frac{\tilde{\Omega}^{(\lambda,\alpha^k,\beta^k)}-\lambda\kappa^{(\alpha^k,\beta^k)}(R^k,D^k)}{2k+3+\lambda\alpha^+ +\sum_{j\in[2:k]}\lambda(2k+3)\alpha_j+\sum_{l\in[k]}2\lambda\alpha_l}\\
&\geq \sup_{\lambda\in[0,1]}\frac{\tilde{\Omega}^{(\lambda,\alpha^{k,*},\beta^{k,*})}-\lambda\kappa^{(\alpha^{k,*},\beta^{k,*})}(R^k,D^k)}{2k+3+\lambda\max_{j\in[k]}\alpha_j^* +\sum_{j\in[2:k]}\lambda(2k+3)\alpha_j^*+\sum_{l\in[k]}2\lambda\alpha_l^*}\\
&\geq \sup_{\lambda\in[0,1]}\frac{\lambda\delta-\frac{\lambda^2\rho}{2}}{2k+9}\label{useresults}\\
&=\frac{\delta^2}{2(2k+9)\rho},
\end{align}
where \eqref{useresults} follows from the results in \eqref{result2touse}, \eqref{result3touse} and the inequality 
\begin{align}
2k+3+\lambda\max_{j\in[k]}\alpha_j^* +\sum_{j\in[2:k]}\lambda(2k+3)\alpha_j^*+\sum_{l\in[k]}2\lambda\alpha_l^*
&\leq 2k+9,
\end{align}
resulting from the constraints that $(\alpha^{k,*},\beta^{k,^*})\in[0,2]^{2k}$ satisfying \eqref{linearconstraints} and $\lambda\in[0,1]$.

\bibliographystyle{IEEEtran}
\bibliography{IEEEfull_lin}

\end{document}

%% file: preamble.tex
\usepackage[mathscr]{eucal}
\usepackage[cmex10]{amsmath}
\usepackage{epsfig,epsf,psfrag}
\usepackage{amssymb,amsmath,amsthm,amsfonts,latexsym}
\usepackage{amsmath,graphicx,bm,xcolor,url,overpic}
\usepackage{fixltx2e}
\usepackage{array}
\usepackage{verbatim}
\usepackage{bm}
\usepackage{algorithmic}
\usepackage{algorithm}
\usepackage{verbatim}
\usepackage{textcomp}
\usepackage{mathrsfs}
\usepackage{epstopdf}

\newcommand{\openone}{\leavevmode\hbox{\small1\normalsize\kern-.33em1}}

\catcode`~=11 \def\UrlSpecials{\do\~{\kern -.15em\lower .7ex\hbox{~}\kern .04em}} \catcode`~=13 

\allowdisplaybreaks[4]

\newcommand{\nn}{\nonumber}

\newcommand{\calA}{\mathcal{A}}
\newcommand{\calB}{\mathcal{B}}

\newcommand{\calF}{\mathcal{F}}
\newcommand{\calG}{\mathcal{G}}

\newcommand{\calM}{\mathcal{M}}

\newcommand{\calP}{\mathcal{P}}
\newcommand{\calQ}{\mathcal{Q}}
\newcommand{\calR}{\mathcal{R}}

\newcommand{\calT}{\mathcal{T}}

\newcommand{\calW}{\mathcal{W}}
\newcommand{\calX}{\mathcal{X}}
\newcommand{\calY}{\mathcal{Y}}
\newcommand{\calZ}{\mathcal{Z}}

\newcommand{\bx}{\mathbf{x}}
\newcommand{\bX}{\mathbf{X}}
\newcommand{\by}{\mathbf{y}}
\newcommand{\bY}{\mathbf{Y}}

\newcommand{\rmc}{\mathrm{c}}

\newcommand{\rme}{\mathrm{e}}

\newcommand{\rmP}{\mathrm{P}}

\newcommand{\rmR}{\mathrm{R}}

\newcommand{\bbN}{\mathbb{N}}

\newcommand{\bbR}{\mathbb{R}}

\DeclareMathAlphabet{\mathbsf}{OT1}{cmss}{bx}{n}
\DeclareMathAlphabet{\mathssf}{OT1}{cmss}{m}{sl}

\DeclareSymbolFont{bsfletters}{OT1}{cmss}{bx}{n}  
\DeclareSymbolFont{ssfletters}{OT1}{cmss}{m}{n}
\DeclareMathSymbol{\bsfGamma}{0}{bsfletters}{'000}
\DeclareMathSymbol{\ssfGamma}{0}{ssfletters}{'000}
\DeclareMathSymbol{\bsfDelta}{0}{bsfletters}{'001}
\DeclareMathSymbol{\ssfDelta}{0}{ssfletters}{'001}
\DeclareMathSymbol{\bsfTheta}{0}{bsfletters}{'002}
\DeclareMathSymbol{\ssfTheta}{0}{ssfletters}{'002}
\DeclareMathSymbol{\bsfLambda}{0}{bsfletters}{'003}
\DeclareMathSymbol{\ssfLambda}{0}{ssfletters}{'003}
\DeclareMathSymbol{\bsfXi}{0}{bsfletters}{'004}
\DeclareMathSymbol{\ssfXi}{0}{ssfletters}{'004}
\DeclareMathSymbol{\bsfPi}{0}{bsfletters}{'005}
\DeclareMathSymbol{\ssfPi}{0}{ssfletters}{'005}
\DeclareMathSymbol{\bsfSigma}{0}{bsfletters}{'006}
\DeclareMathSymbol{\ssfSigma}{0}{ssfletters}{'006}
\DeclareMathSymbol{\bsfUpsilon}{0}{bsfletters}{'007}
\DeclareMathSymbol{\ssfUpsilon}{0}{ssfletters}{'007}
\DeclareMathSymbol{\bsfPhi}{0}{bsfletters}{'010}
\DeclareMathSymbol{\ssfPhi}{0}{ssfletters}{'010}
\DeclareMathSymbol{\bsfPsi}{0}{bsfletters}{'011}
\DeclareMathSymbol{\ssfPsi}{0}{ssfletters}{'011}
\DeclareMathSymbol{\bsfOmega}{0}{bsfletters}{'012}
\DeclareMathSymbol{\ssfOmega}{0}{ssfletters}{'012}

\newcommand{\tilC}{\tilde{C}}

\newcommand{\tilF}{\tilde{F}}

\newcommand{\hatx}{\hat{x}}
\newcommand{\hatX}{\hat{X}}

\newcommand{\bard}{\bar{d}}

\newtheorem{theorem}{Theorem} 
\newtheorem{lemma}[theorem]{Lemma}

\newtheorem{definition}{Definition}